\documentclass[usenatbib]{mn2e}
\usepackage{setspace}
\usepackage{epsf}
\usepackage{graphicx}
\usepackage{color}
\usepackage{txfonts}
\usepackage{tabularx}
\usepackage{here}
\usepackage{float}

\newcommand{\sii}{[S\,{\sc ii}]}

\usepackage{setspace}

\newcommand{\apj}{ApJ}
\newcommand{\aj}{AJ}

\newcommand{\apjs}{ApJ Supplements}
\newcommand{\aap}{A$\&$A}
\newcommand{\mnras}{MNRAS}
\newcommand{\aaps}{A\&AS}
\newcommand{\apjl}{ApJ Letters}
\newcommand{\nat}{Nature}
\newcommand{\pasj}{PASJ}
\newcommand{\pasp}{PASP}
\newcolumntype{Y}{>{\centering\arraybackslash}X} 
\newcolumntype{Z}{>{\raggedleft\arraybackslash}X}

\title[C$_{60}$ in Galactic PNe]{Physical Properties of Fullerene-containing Galactic Planetary Nebulae}

\author[Otsuka et al.]{Masaaki
Otsuka$^{1}$\thanks{E-mail:otsuka@asiaa.sinica.edu.tw},
F.~Kemper$^{1}$, 
J.~Cami$^{2,3}$, 
E.~Peeters$^{2,3}$,
J.~Bernard-Salas$^{4,5}$
\\
$^{1}${Institute of Astronomy and Astrophysics, Academia Sinica, 
11F of Astronomy-Mathematics Building, AS/NTU. No.1, Sec. 4, Roosevelt
Rd, Taipei 10617, Taiwan, R.O.C.}\\
$^{2}${Department of Physics and Astronomy, The University of Western
  Ontario, London, ON N6A 3K7, Canada}\\
$^{3}${SETI Institute, 189 Bernardo Ave, Suite 100, Mountain View, CA
  94043, USA}\\
$^{4}${Institut d'Astrophysique Spatiale,
  CNRS/Universit\'{e} Paris-Sud 11, 91405 Orsay, France}\\
$^{5}${current address: Department of Physical Sciences, The
Open University, Milton Keynes, MK7 6AA, UK}
}

\begin{document}

\date{}

\pagerange{\pageref{firstpage}--\pageref{lastpage}} \pubyear{2013}

\maketitle

\label{firstpage}

\begin{abstract}
  We searched the \emph{Spitzer} Space Telescope data archive for
  Galactic planetary nebulae (PNe), that show the
  characteristic 17.4 and 18.9 $\mu$m features due to C$_{60}$, also
  known as buckminsterfullerene. Out of 338 objects
  with \emph{Spitzer}/IRS data, we found eleven C$_{60}$-containing PNe,
  six of which (Hen2-68, IC2501, K3-62, M1-6, M1-9, and SaSt2-3) are new
  detections, not known to contain C$_{60}$ prior to this work. The
  strongest 17.4 and 18.9 $\mu$m C$_{60}$ features are seen in Tc 1
  and SaSt 2-3, and these two sources also prominently show the
  C$_{60}$ resonances at 7.0 and 8.5 $\mu$m. 
In the other nine sources, the 7.0 and 8.5 $\mu$m features due to C$_{60}$ are much
  weaker. We analyzed the spectra, along with ancillary data, using 
the photo-ionization code {\sc CLOUDY} to establish the atomic line
  fluxes, and determine the properties of the radiation field, as set
  by the effective temperature of the central star. In addition, we
  measured the infrared spectral features due to dust grains. 
We find that the Polycyclic Aromatic Hydrocarbon (PAH) 
profile over 6-9 $\mu$m in these C$_{60}$-bearing carbon-rich
 PNe is of the more chemically-processed class A. 
The intensity ratio of 3.3 $\mu$m to 11.3 $\mu$m PAH indicates that the
 number 
of C-atoms per PAH in C$_{60}$-containing PNe is small compared to
that in non-C$_{60}$ PNe. The \emph{Spitzer} spectra also show broad dust features
  around 11 and 30 $\mu$m. Analysis of the 30-$\mu$m feature shows
  that it is strongly correlated with the continuum, and we propose
  that a single carbon-based carrier is responsible for both the
  continuum and the feature. The strength of the 11-$\mu$m feature is
  correlated to the temperature of the dust, suggesting that it is at
  least partially due to a solid-state carrier. The chemical abundances
 of C$_{60}$-containing PNe can be explained by AGB nucleosynthesis 
models for initially 1.5-2.5 M$_{\odot}$ stars with $Z$=0.004. 
We plotted the locations of C$_{60}$-containing PNe on a
 face-on map of the Milky Way and we found that most of these PNe are
 outside the solar circle, consistent with low metallicity values. Their
 metallicity suggests that the progenitors are an older population. 
\end{abstract}

\section{Introduction}

Since their laboratory discovery in 1985 \citep{Kroto:1985aa},
fullerenes, particularly C$_{60}$, have drawn considerable interest
from astrochemists looking for them in interstellar conditions.
Fullerenes are extremely stable and easily form in laboratories on the
earth, so it had been thought that they should exist in interstellar
space. However, the first confirmed detection of cosmic fullerenes was
only recently reported by \citet{Cami_2010} in the C-rich planetary
nebula (PN) Tc1. Since then, a plethora of papers reported the discovery
of cosmic fullerenes in Galactic
(\citealt{2010ApJ...724L..39G,2011ApJ...737L..30G,Otsuka_2013,2011ApJ...730..126Z})
and Large/Small Magellanic PNe
(\citealt{2010ApJ...724L..39G,2011ApJ...737L..30G}), Galactic
asymptotic giant branch (AGB) stars (\citealt{2011A&A...536A..54G}),
H-poor R Coronae Borealis (R CrB) stars
(\citealt{Gar_2011,2011AJ....142...54C}), reflection nebulae
(\citealt{2010ApJ...722L..54S,Peeters:2012aa}), a binary XX Oph star
(\citealt{2012MNRAS.421L..92E}), and YSOs
(\citealt{2012MNRAS.421.3277R}).  While 
the number of fullerene detections is increasing, the formation 
process and the excitation mechanism of fullerenes in evolved stars 
are the subject of active research. For instance, a recent work 
of \citet{Micelotta:2012aa} details a possible formation mechanism, 
while \citet{Jero:C60excitation} provides an observational/theoretical 
study of the fullerene excitation conditions. In order to allow for 
further investigation of the formation and  excitation mechanisms, 
it is of crucial importance to characterize the observational and 
physical properties of the environments in which fullerenes are found. 
PNe provide an excellent opportunity to do this.

PNe represent the final stage in the evolution of initially 1-8
M$_{\odot}$ stars. At the end of its life, such a star evolves
first into a red giant branch star, then an AGB star, finally a PN. In
the early AGB stage, dust grains are formed above the surface of the
star. Dust grains receive radiation pressure from the central star 
and move radially outward. The grains drag the H-rich envelope;
as a result, dramatic mass-loss begins. Dust controls not only the
mass-loss history of the stars but also the evolution of the ISM in
galaxies. To understand the stellar evolution and the ISM evolution in
galaxies, it is important to understand the amount, composition,
formation process of the dust grains produced by dying stars.

According to stellar evolution theory, stars with a main-sequence mass
$\sim$1.2-3.5 M$_{\odot}$ can efficiently synthesize C by
nucleosynthesis taking place in the He-rich intershell and the third
dredge-up during the late AGB phase, so that these stars evolve into
C-rich AGB stars/PNe. Several species of C-rich dust can form in the
ejected nebula. Using infrared space telescopes such as \emph{ISO},
\emph{Spitzer}, and \emph{AKARI}, the presence of these dust species
was confirmed using the characteristic emission bands in mid-infrared
spectra in the range from $\sim$3 to $\sim$40 $\mu$m; for example,
amorphous carbon, polycyclic hydrocarbon (PAH), the 11 $\mu$m silicon
carbide (SiC) band, and the unidentified 30 $\mu$m feature are detected.  C-rich
evolved stars are important producers of C based dust grains in galaxies.

In order to address the outstanding questions on the formation and the excitation
mechanism of C$_{60}$ in evolved stars, in particular, PNe, 
it is necessary to find more fullerene containing PNe in the
entire \emph{Spitzer} data archive, measure the C$_{60}$ band as accurately
as possible, and investigate the properties of dust, ionized gas, 
and central stars in these C$_{60}$-containing PNe.

In this paper, we present results that are based on {\em all}
Spitzer/IRS observations of Galactic PNe (338 objects in total).  
Our search resulted in six new detections: Hen2-68, IC2501, 
K3-62, M1-6, M1-9, and SaSt2-3. In
addition, we found the 17.4 and 18.9 $\mu$m C$_{60}$ bands in the
\emph{Spitzer}/IRS SH spectrum of IC418; \citet{Morisset_2012}
reported the C$_{60}$ detection in \emph{ISO} spectra. 
We compare the \emph{Spitzer}, \emph{AKARI}, and
\emph{ISO} mid-IR spectra of these objects and previously known
fullerene-containing PNe M1-11 \citep{Otsuka_2013}, M1-12, M1-20
\citep{2010ApJ...724L..39G}, and Tc1 \citep{Cami_2010} to the
properties of the central stars and the other dust features such as
the 3.3 and the 6-9 $\mu$m PAH bands, the broad 11 $\mu$m and 30
$\mu$m features, which are frequently seen in C$_{60}$ PNe. 
Moreover, we construct photo-ionization models using
{\sc Cloudy} (\citealt{1998PASP..110..761F}) to derive the physical
properties of the central stars and the nebulae of the eleven PNe 
in our sample.

\section{Observations \& Data Reduction}
\label{Sect:Observations}

\begin{table*} 
\caption{\emph{Spitzer}, \emph{ISO}, and \emph{AKARI} observing log.\label{Table:obs-log1}}
\begin{tabularx}{\textwidth}{@{}lcYlclrcYcclr@{}} 
\hline
\hline
{}&
{}&
\multicolumn{4}{c}{Spitzer/ISO}&
{}&
\multicolumn{4}{c}{AKARI}\\
\cline{3-6}
\cline{8-11}
{Nebula}  &
{PN G} &
{AOR Key/} &
{Obs. Date}&
{PI}&
{SNR}&
{}&
{OBSID}&
{Obs. Date}&
{PI}&
{SNR}\\
{}&
{lll.l$\pm$bb.b}&
{TDT}&
{yy-mm-dd}&
{}&
{}&
{}&
{}&
{yy-mm-dd}\\
\hline
Hen2-68 &294.9--04.3 &25852417&09-04-03&L. Stanghellini&138&&3460047&09-01-26&T. Onaka&26\\   
IC418   &215.2--24.2 &4909568 &04-10-01&D. Cruikshank&105&&Saturated\\
        &            &82901301$^{\rm a}$&98-02-22&D. Beintema&87&\\
        &            &86801205$^{\rm b}$&98-04-01&M. Barlow&24&\\
IC2501  &281.0--05.6 &4905728 &04-05-13&D. Cruikshank&747&&1740559&08-07-01&Y. Nakada&54\\
K3-62   &095.2+00.7  &25847808&08-08-06&L. Stanghellini&208& &3460033&08-12-18&T. Onaka&17\\ 
M1-6    &211.2--03.5 &21953024&08-12-07&J.~Bernard-Salas&318&&3460036&09-03-29&T. Onaka&44\\
M1-9     &212.0+04.3 &2194560&08-04-27&J.~Bernard-Salas&83 &&\\
M1-11   &232.8--04.7 &19903232&06-11-09&H.~Dinnerstein&211&&3460037&09-04-11&T. Onaka&52\\
M1-12   &235.3--03.9 &25849856&08-11-30&L. Stanghellini&138&&3460038&09-04-15&T. Onaka&39\\
M1-20   &006.1+08.3  &11315712&05-04-19&M. Bobrowsky&38&&\\
SaSt2-3 &232.0+05.7  &21958144&08-04-29&J.~Bernard-Salas&16&&\\
Tc1     &345.2--08.8 &11321600&05-03-21&M. Bobrowsky&57 &&\\
\hline
\end{tabularx}
\raggedright
\begin{tabular}{l}
$^{a}${SWS mode}\\
$^{b}${LWS mode}
\end{tabular}
\end{table*}

Our aim is to study in detail the properties of the gas and dust
  in Galactic fullerene-containing PNe, to provide constraints on
  C$_{60}$ formation and excitation, and to find clues to the
  formation and processing of carbonaceous dust. Mid-infrared
  observations at high spectral resolution are crucial for this, and
  we thus selected our sample based on the availability of the
  \emph{Spitzer}/IRS SL (5.2-14.5 $\mu$m), SH (9.9-19.6 $\mu$m), and LH
  (18.7-37.2 $\mu$m) spectra and \emph{ISO} spectra. 
  Where available, we have also included data at shorter
  wavelengths from \emph{AKARI}/IRC 2.5-5 $\mu$m spectroscopy, 
  as well as optical and infrared images from the Hubble Space Telescope
  (\emph{HST}), the 8.2-m Gemini telescopes, and the ESO/NTT 3.6-m telescope 
to get a better understanding of the properties of gas and dust.

The observation logs for \emph{Spitzer}, \emph{ISO}, and \emph{AKARI} are
summarized in Table \ref{Table:obs-log1}. The second column is the
position of each PN in the Galactic coordinates. The information of
\emph{Spitzer} and \emph{ISO} observations is summarized in the third
to sixth columns.  The signal-to-noise ratios (SNR) are measured
between 18.5 and 19.5 $\mu$m for \emph{Spitzer} and \emph{ISO}/SWS and
between 148 and 154 $\mu$m for \emph{ISO}/LWS spectra, respectively.
The information of \emph{AKARI}/IRC observations are in the remaining 
columns. The SNRs are measured between 3.3 and 3.6 $\mu$m.

\subsection{Spitzer-IRS and ISO spectra}

\begin{table}
\caption{The flux density at the \emph{WISE} band 4 and the adopted 
scaling factor for \emph{Spitzer}/IRS and \emph{ISO}. \label{Table:scaling}}
\begin{tabularx}{\columnwidth}{@{}lYcYc@{}}
\hline\hline
{Nebula}&
{$F_{\nu}$(\emph{WISE}4)}&
{Scaling Factor}\\
{}&
{(Jy)}&
{}\\
\hline
Hen2-68 &4.34  &1.288\\
IC418   &150.33&1.203\\
IC2501  &19.40 &2.519$^{a}$\\
K3-62   &9.56  &1.215\\
M1-6    &9.10  &1.251\\
M1-9    &1.36  &1.081\\ 
M1-11   &52.51 &1.181\\
M1-12   &9.13  &1.224\\
M1-20   &3.17  &1.257\\
SaSt2-3 &0.242 &1.147\\
Tc1     &10.08 &3.858\\
\hline
\end{tabularx}
\raggedright
\begin{tabular}{l}
$^{a}${determined using the 2MASS $Ks$ image. See text for detail.}\\
\end{tabular}
\end{table}

We searched the entire \emph{Spitzer} Infrared Spectrograph (IRS,
\citealt{Houck:2004aa}) archive for PNe with observations in the SL (5.2-14.5
$\mu$m), SH (9.9-19.6 $\mu$m), and LH (18.7-37.2 $\mu$m) modules.  In this search, we found six new
fullerene-containing PNe, namely, Hen2-68, IC2501, K3-62, M1-6, M1-9, 
and SaSt2-3.  \citet{Morisset_2012} reported that IC418 shows two
resonances at 17.4 and 18.9 $\mu$m in the \emph{ISO} spectra 
due to C$_{60}$.  We confirmed both of these C$_{60}$
lines in the \emph{Spitzer} spectrum of IC418, which was not
previously analyzed. We combine this data set with \emph{Spitzer}/IRS
spectra of M1-11, M1-12, M1-20, and Tc1, which are known to show the
17.4 and 18.9 $\mu$m C$_{60}$ resonances.
We did not include the C$_{60}$ PN K3-54 \citep{2010ApJ...724L..39G}, 
because only \emph{Spitzer}/IRS SL and LL spectra are available.

To reduce the \emph{Spitzer}/IRS data, 
we used the reduction package SMART v.8.2.5 provided by the IRS 
Team at Cornell University (\citealt{Higdon_2004}) and IRSCLEAN 
provided by the \emph{Spitzer} Science Center. With IRSCLEAN, we removed
the bad pixels. For the SH and LH spectra, we subtracted the 
sky background using off-set spectra taken in the same program as the
on-source spectra, if available. We scaled the flux density of the SL
spectra up to,  
and the LH spectra down to, the SH spectra using the overlapping wavelength regions. 
Finally, we scaled the flux density of these combined spectra to the Wide-field 
Infrared Survey Explorer 
(\emph{WISE}) band 4 photometric value ($\lambda_{\rm c}$=22 $\mu$m), except for IC2501. 
The flux density at the \emph{WISE} band 4 and the adopting scaling factor matching this 
band for each object are listed in Table \ref{Table:scaling}.

For IC2501, we estimated the light loss using the Two Micron All-Sky
Survey (2MASS) $Ks$ band image. By measuring the pixel values of the entire
nebula and those inside the area of the SH slit, we estimated that $\sim$39.7$\%$ of 
the light from IC2501 was included in the SH entrance of \emph{Spitzer}. Thus, we scaled the flux
density of the spectrum by a factor of 2.519.

For IC418, we scaled the flux density of the \emph{Spitzer} SH spectrum 
up to match that of \emph{ISO}. The spectral resolution 
of the \emph{ISO} spectrum was reduced to $\sim$600 to match that 
of \emph{Spitzer}/SH by the Gaussian convolution method. Then, we combined 
this into a single 2.5-170 $\mu$m spectrum. Finally, we scaled the 
flux density of this spectrum to the \emph{WISE} band 
4 photometry with a scaling factor of 1.203 (See Table \ref{Table:scaling}). 
In the 10-20 $\mu$m wavelength range we adopted the
\emph{Spitzer}/IRS data, and outside this range we used the \emph{ISO} spectrum.

\subsection{AKARI/IRC spectra \label{akari_pah}}
Checking for the presence of the 3.3 $\mu$m PAH feature 
is necessary to estimate the extent by which the 8.5 $\mu$m C$_{60}$ band is
contaminated by the 8.6 $\mu$m PAH feature.
Using the intensity ratio of 3.3 $\mu$m to 11.3 $\mu$m PAH bands, we can
estimate the number of C atoms of each PAH grain.

To that end, we analyzed the 2.5-5.5 $\mu$m spectra of Hen2-68,
IC2501, K3-62, M1-6, M1-11 and M1-12 taken with the Infrared Camera
spectrograph (IRC; \citealt{Onaka_2007}) on board of the \emph{AKARI}
satellite (\citealt{Murakami_2007}).  The data were obtained as part
of two different mission programs, AGBGA (IC2501, PI: Y. Nakada) and
PNSPEC (the remainder, PI: T. Onaka).  The 1$'$$\times$1$'$ observing window is large
enough to include the optical nebulae of these PNe as demonstrated in
Section \ref{S:image}, therefore, we assume that the loss of the light
is negligibly small and we did not perform any correction in flux
density.  For the data reduction, we used the IRC Spectroscopy Toolkit
for the Phase 3 data version.

\subsection{Optical and mid-infrared images of C$_{60}$-containing
  PNe \label{S:image}} 
To check the apparent size of the nebulae, the nebular shape, and the
slit-position in the \emph{Spitzer} observations, we analyzed the
archival optical and mid-infrared images from \emph{HST}, 
Gemini 8.2-m, and ESO/NTT 3.6-m telescopes.
These images are necessary to set the outer radii of each PN 
in {\sc Cloudy} photo-ionization models (See Section \ref{S:cloudy}).

We downloaded the archival data of IC2501 and K3-62 from the Canadian
Astronomy Data Centre (CADC) taken by the Gemini telescopes 
with the mid-infrared imager and spectrograph T-ReCS for IC2501
and Michelle for K3-62. These data were taken through the observing programs
GS-2004B-DD-8 (PI: K.Volk) for IC2501 and GN-2005A-C-4 (PI: S.Kwok) for
K3-62. Both images were taken using the Si-5 filter 
($\lambda_{c}$=11.6 $\mu$m). We reduced the data of IC2501 and K3-62 using
NOAO/IRAF\footnote[6]{IRAF is distributed by the National Optical Astronomy
  Observatories, which are operated by the Association of Universities
  for Research in Astronomy (AURA), Inc., under a cooperative
  agreement with the National Science Foundation.} in a standard
  manner. The archival \emph{HST} data of Hen2-68 (WFC3/F507N), IC418, 
M1-6, M1-11, M1-12, and M1-20 (all WFPC2/F656N) were also downloaded 
from the CADC and reduced using the standard pipeline with 
STSDAS/MultiDrizzle to obtain higher resolution images. 
The NTT/EFOSC data of Tc1 taken through the program
64.H-0557(A) (PI: T.Rauch) were downloaded from the ESO archive
center, and we reduced the $R$-band image using IRAF. The 
instrumental distortion was corrected with the XYXYMATCH, GEOMAP, and 
GEOTRAN IRAF routines, and we aligned the positions of stars in the detector.

\begin{figure*}
\centering
\includegraphics[clip,scale=0.9]{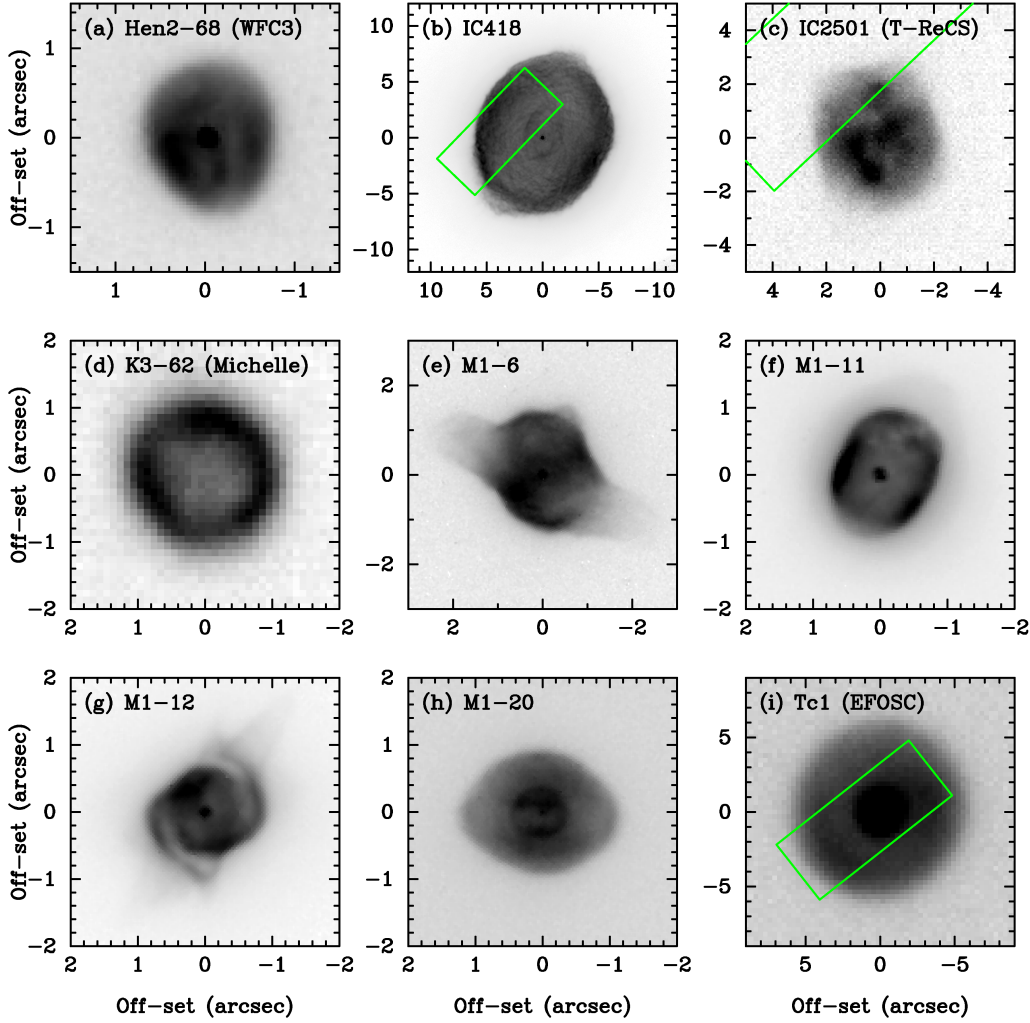}
\caption{\label{image} 
Imaging of the objects in our sample. North is up and east is left. 
In IC418, IC2501, and Tc1, the position and the 
size of \emph{Spitzer}/IRS SH slit are indicated 
by the green boxes. The slit size of the SH module is larger than the 
apparent size for the other PNe.} 
\end{figure*}

The reduced images are presented in Fig. \ref{image}. Since our attention
is mainly towards the regions where the 17.4 and 18.9 $\mu$m C$_{60}$ bands are detected, we
also present the slit position and its size for the \emph{Spitzer}/IRS SH
module (slit dimension: 4.7$''$$\times$11.3$''$), indicated by the green
boxes for IC418, IC2501, and Tc1. For the other PNe which do not show
the green boxes, the slit size is
larger than the apparent size of the nebulae, and thus, the SH
measurements have collected all the light emitted by the nebulae. 
We do not show the images of M1-9 and SaSt2-3, because
there are no high-resolution images available. 
We were able to measure the nebular size of these two objects using the 
2MASS $J$-band images which show that M1-9 and SaSt2-3 have round shaped
nebulae with $\sim$4$''$ and $\sim$2$''$ radii, respectively.

There are no common characteristics in the morphology of the nebula shape. 
The C$_{60}$-containing PNe show different nebular shapes; 
ring (K3-62), round (Hen2-68, IC2501, Tc1), elliptical (IC418, 
M1-12, M1-20), bipolar (M1-6), and multi-polar (M1-12). Moreover, 
blobs are seen in  Hen2-68 ($\sim$1$''$ SE from the
center of the nebula), in IC2501 ($\sim$2$''$ NW and SE), and in Tc1 
($\sim$5$''$ SE). We confirm that Tc1 has a faint optical nebula which extends 
up to $\sim$25$''$ (not presented here; See Fig. 6b of \citealt{Williams:2008aa}).

The Si-5 narrow band images of IC2501 and K3-62 cover the wavelength from
$\sim$10.8 to $\sim$12.5 $\mu$m. Both PNe show 
the broad 11 $\mu$m feature in their \emph{Spitzer}/IRS spectra, 
as we present below. The carrier of this broad dust feature in 
IC2501 exists in the nebula as blobs, while that in K3-62 is smoothly 
distributed and its density peak is $\sim$0.8$''$ far away from 
the central star.

\section{Properties of the central stars and the nebulae}
\label{S:cloudy}

\begin{table*}
\caption{Properties of the central stars and the
 nebulae derived photo-ionization models. \label{Table:properties}}
\begin{tabularx}{\textwidth}{@{}XcYccccccc@{}}
\hline\hline
{Nebula} & 
{$T_{\rm eff}$} & 
{$\log$~($L_{\ast}$/L$_{\odot}$)} & 
{$\log$~$g$} &
 {Type of} &
{P-Cygni}&
{$d$} & 
{C/N/O/Ne/S/Ar} & 
{$F$([Ar\,{\sc ii}] 6.99 $\mu$m)/} \\
{} & 
{(K)} &  
{}& 
{(cm s$^{-2}$)} &
{central star} &
{}& 
{(kpc)} & 
{(Log~X/H+12)}&
{$F$([Ar\,{\sc iii}] 8.99
 $\mu$m)}\\
\hline
Hen2-68 & 40\,800 & ? & ? & ? &?&4.90 & ? & 0.65 \\ 
IC418 & 36\,700 & 3.88 & 3.55 &H-rich&y& 1.26 & 8.90/8.00/8.60/8.00/6.65/6.20 & 1.0 \\ 
IC2501 & 51\,650 & 3.17 & 4.75 &wels?&y& 2.20 & 8.77/7.88/8.58/7.94/6.56/6.13 & 0.03 \\ 
K3-62 & 45\,000 & ? & ? &?&?& 6.00 & ? & 0.09 \\ 
M1-6 & 34\,960 & 3.31 & 3.50 &[WC10-11]&? &4.10 & 8.76/7.91/8.60/8.05/6.49/5.95 & 0.18 \\ 
M1-9 & 43\,500 & 3.41 & 4.50 &?&? &9.0 & 8.81/7.69/8.34/7.68/6.25/5.89 & 0.04\\
M1-11 & 31\,830 & 3.67 & 3.30 &[WC10-11]&? &2.10 & 8.49/7.89/8.30/7.67/6.15/5.95 & 5.20\\ 
M1-12 & 31\,660 & 3.54 & 3.40 &[WC10-11]&? &3.90 & 8.74/7.75/8.50/7.80/6.48/5.81 & 1.99\\ 
M1-20 & 45\,880 & 3.59 & 4.75 &wels&?& 7.30 & 8.70/7.85/8.58/7.85/6.76/5.94 & 0.02\\ 
SaSt2-3 & 29\,750 & 2.80 & 3.30 &OB. &?&6.00 & 8.72/7.49/8.23/7.68/6.17/5.93 & 4.71 \\
Tc1 & 34\,060 & 3.40 & 3.30 & Of(H)&y &3.00 & 8.71/7.83/8.61/7.98/6.58/6.07 & 0.10\\
\hline
\end{tabularx}
\end{table*}

Before studying the IR spectra of our sample, we 
infer the overall physical properties of the central stars as 
well as the nebular conditions. 

To derive physical parameters of the central stars and the nebulae, we
run photo-ionization models for each PN except for Hen2-68, K3-62,
IC418, and M1-11 using {\sc Cloudy} (\citealt{1998PASP..110..761F}). 
We refer to the results by \citet{Morisset:2009aa} for 
IC418 and \citet{Otsuka_2013} for M1-11. Since there is no information 
on gas emission line fluxes, except for H$\beta$, the effective temperature
$T_{\rm eff}$, and the distance $d$, we could not model 
Hen2-68 and K3-62 using {\sc Cloudy}. The effective temperatures 
($T_{\rm eff}$) of Hen2-68 and K3-62 are
from \citet{Preite-Martinez:1989aa} and \citet{Kondratyeva:2003aa},
respectively. These models allow us to estimate the 
contribution from [Ar\,{\sc ii}] 6.99 $\mu$m line to the 7.0 $\mu$m C$_{60}$ band.

\subsection{{\sc Cloudy} modeling approach}
In the {\sc Cloudy} calculations, we used Tlusty's non-LTE theoretical
atmosphere models for O stars (\citealt{Hubeny:1995aa}).
The surface gravities $\log{g}$ of hot white dwarfs ($T_{\rm
eff}$$\gtrsim$50\,000 K) are sometimes $\gtrsim$6. If our sample
contains such hot white dwarfs, they should have highly ionized nebulae showing 
high ionization potential lines such as He\,{\sc ii} lines and [O\,{\sc iv}] 25.9
$\mu$m. However, our sample does not show such lines in the nebulae. 
Since the absorption line analysis of the central star of IC418 shows
that the
$\log{g}$$\sim$4 \citep{Morisset:2009aa,1989ApJ...345..327P}, we
assume that the $\log{g}$ of other objects in our sample are in the range between 
$\sim$3 and $\sim$5. We used a series of theoretical atmosphere models with 
$T_{\rm eff}$ in the range from 27\,500 to 55\,000 K 
and the $\log{g}$ in the range from of 3.0 to 
4.75 to describe the SED of the central star. We referred to
\citet{1993ApJS...88..137Z} for the $\log{g}$ of M1-6, M1-9, 
M1-12, M1-20, and Tc1 as the first estimate. For SaSt2-3, we used the same $\log{g}$ as for
Tc1.

We adopted a constant hydrogen (H) density for all PNe. 
We fixed the outer radius at the values measured using the images shown in 
Fig. \ref{image} and varied the 
inner radius with the H density to match the observed de-reddened 
H$\beta$ flux of the whole nebula measured by \citet{Dopita:1997aa} for SaSt2-3 and 
\citet{1992A&AS...94..399C} for the others. The distances to each PN
estimated by \citet{Tajitsu:1998aa} are adopted, except for SaSt2-3. 
For this source, we adopted the upper value of \citet{Pereira:2007aa},
namely 6 kpc.

For the gas-phase elemental abundances X/H, we adopted the 
observed values by prior works as a first guess and 
varied these to match the observed flux densities of over 30 emission 
lines of He/C/N/O/Ne/S/Cl/Ar in UV to mid-IR wavelength, except 
for SaSt2-3 (14 lines). We adopted the line-ratios measured by
\citet{Sharpee:2007aa} for IC2501 and \citet{Henry:2010aa} for M1-6,
M1-9, and M1-12, by \citet{2007MNRAS.381..669W} for M1-20, 
\citet{Pereira:2007aa} for SaSt2-3, and compiled by
\citet{Pottasch:2011aa} 
for Tc1. The emission-line fluxes in the mid-IR wavelength range 
were measured by us for all objects in our sample.
We did not include the [Ar\,{\sc ii}] 6.99 $\mu$m as an Ar
abundance constraint. 

Thus, we considered line fluxes from He\,{\sc i}, C\,{\sc ii}, C\,{\sc
iii}$]$, [N\,{\sc ii}], [O\,{\sc i,ii,iii}], [Ne\,{\sc ii,iii}], [S\,{\sc
ii,iii,iv}], [Cl\,{\sc iii}], and [Ar\,{\sc iii,iv}].

There are no UV spectra of M1-6, M1-9, M1-12, M1-20, and SaSt2-3 
required to determine the C abundances using 
collisionally excited lines such as C~{\sc iii}] 1906/09 {\AA}.
For the C abundances of these PNe we set the value according to
\begin{equation}
{\rm [C/Ar]=-0.89\times[Ar/H]+0.49,}
\end{equation}
\noindent which is established among 115 Milky Way PNe and 
the C abundances are estimated using collisionally 
excited C emission lines in the UV wavelength (Otsuka in preparation). 
We updated 
collisional strengths and transition probabilities 
of of C\,{\sc iii}], [O\,{\sc ii,iii}], 
[N\,{\sc ii}], [Ne\,{\sc ii,iii,iv}], [S\,{\sc ii,iii}], [Cl\,{\sc ii,iii}], and 
[Ar\,{\sc ii,iii,iv}] lines, which are the same data listed in Table 7 of \citet{Otsuka:2010aa}. Reliable 
dielectric recombination (DR) rate measurements do not exist for low 
stages of ionization of S at photo-ionization temperatures ({\sc 
Cloudy} manual), thus we adopted the scaled DR rate of oxygen for the 
sulfur line calculations, to match the observed {\sii}.

\subsection{Results of the {\sc Cloudy} modeling and comparison with
  non C$_{60}$-containing PNe} 
The results of the best model fits are listed in Table
\ref{Table:properties}. The second to fourth columns show the
effective temperatures, the luminosities $L_{\ast}$ of the central
stars, and surface gravities, respectively. The uncertainties in the 
$T_{\rm eff}$ and $\log$~$g$ are within 1000 K and 0.2 cm s$^{-2}$,
respectively. 
Since the distances to each PN have a large uncertainty, typically a
factor of $\sim$1.5-2, the uncertainty in 
$\log$($L_{\ast}$/L$_{\odot}$) is $\sim$0.35-0.6.
The fifth column is the type of the central star, from
\citet{Weidmann:2011ab,Weidmann:2011aa}. The predicted $F$([Ar\,{\sc
ii}] 6.99 $\mu$m)/$F$([Ar\,{\sc iii}] 8.99 $\mu$m) ratios, which are
used in the 7.0 $\mu$m C$_{60}$ flux measurements (See Section
\ref{S:flux-measurement} and Appendix), are listed in the last column.

The gas abundances derived by the models are in the eighth column.
The uncertainties in the derived gas abundances are within 0.1 dex. The
difference between the observed and our model predicted gas abundances is within 0.1-0.2 dex.
All objects are C-rich (i.e., the C/O ratio $>$1) and the elemental
composition is very similar within the sample, suggesting that these PNe
have evolved from similar progenitor stars. 
As we mentioned above, 
the C abundances in M1-6, M1-9, M1-12, and M1-20, and SaSt2-3 are not 
determined using collisionally excited C lines yet. 
In order to accurately measure the C/O ratio, we need to detect 
collisionally excited C lines or C and O recombination lines in these 
C$_{60}$-containing PNe, in the future. 
The average Ar abundance,
which indicates the initial metallicity of the progenitors, among 9 of
the objects in our sample is 5.99 (Ar=6.55;
\citealt{Lodders:2003aa}), corresponding to the metallicity of
$Z\sim$0.005 in Ar (0.27 Z$_{\odot}$), which is close to the typical
metallicity of the Small Magellanic Cloud. Plotting the estimated $L_{\ast}$
and $T_{\rm eff}$ on the He-burning tracks from
\citet{1994ApJS...92..125V} with $Z$=0.004 shows that the initial mass
of the objects is $\sim$1.5-2.5 M$_{\odot}$. The chemical abundances of
our C$_{60}$-containing PNe can be explained by the AGB nucleosynthesis
models for the initially 1.5-2.5 M$_{\odot}$ stars with $Z$=0.004 
by \citet{Karakas:2010aa} (Table \ref{agbabund}).

In the last line of Table \ref{agbabund}, we list the average abundances
of non C$_{60}$-containing Galactic C-rich PNe taken from
\citet{Pottasch:2006aa}. The abundances of C$_{60}$ PNe are not very
different from those of these non C$_{60}$-containing PNe, except for
N. This might be due to the two $\alpha$ capturing by $^{14}$N, because
the Ne abundances in C$_{60}$-containing PNe are slightly larger than the
model predictions. The low Ar abundance in C$_{60}$-containing PNe 
(5.81-6.20) suggests that the progenitors are an older population.

\subsection{Presence of the stellar wind}
We checked the presence of P-Cygni profiles in the Ly$\alpha$ 1215
{\AA} line in the archived Far Ultraviolet Spectroscopic Explorer (\emph{FUSE})
and International Ultraviolet Explorer (\emph{IUE}) spectra, as listed in the sixth
column. We found that IC418, IC2501, and Tc1 indeed show 
the P-Cygni profile. There are no data for
Hen2-68, K3-62, SaSt2-3, M1-6 and M1-12 and no
data for M1-11 with enough SNR to check for the presence of a P-Cygni
Ly$\alpha$ profile. However, since M1-6, M1-11, M1-12, and
M1-20 have Wolf-Rayet like central stars ([WC10,11]; weak emission
line stars, {\it wels}), which show the broad C\,{\sc iii,iv} lines, there has
to be a strong wind (e.g., \citealt{Acker:2003aa}).

\begin{table}
\caption{Comparison of the predicted abundances for different initial 
mass stars with $Z$=0.004 by \citet{Karakas:2010aa} and the observed abundances. \label{agbabund}} 
\begin{tabularx}{\columnwidth}{@{}l@{\hspace{-1.pt}}ccccc@{}}
\hline\hline
initial    &C&N&O&Ne&S\\
mass\\
\hline
1.50 M$_{\odot}$&8.46&7.65&8.23&7.42&6.70\\
1.75 M$_{\odot}$&8.80&7.70&8.25&7.51&6.70\\
1.90 M$_{\odot}$&8.94&7.71&8.25&7.61&6.70\\
2.10 M$_{\odot}$&9.00&7.68&8.25&7.67&6.70\\
2.25 M$_{\odot}$&9.26&7.72&8.27&8.00&6.72\\
2.50 M$_{\odot}$&9.38&7.76&8.27&8.20&6.73\\
\hline
C$_{60}$ PNe &8.49-8.90&7.49-8.00&8.23-8.61&7.67-8.05&6.15-6.76\\
\hline
non-C$_{60}$ &8.85     &8.23     &8.56     &8.06     &6.85     \\
C-rich PNe$^{a}$\\
\hline
\end{tabularx}
\raggedright
$^{a}${The averaged value between 11 C-rich Galactic PNe listed in 
Table A1 of \citet{Pottasch:2006aa}. The averaged Ar abundance is 6.42.}
\end{table}

\section{The IR spectra of galactic C$_{60}$ PNe}
\label{Sect:IRspectra}

\subsection{Overview: Spectral Features and Variations in 5-36 $\mu$m}

\begin{figure}
\centering
\includegraphics[width=\columnwidth]{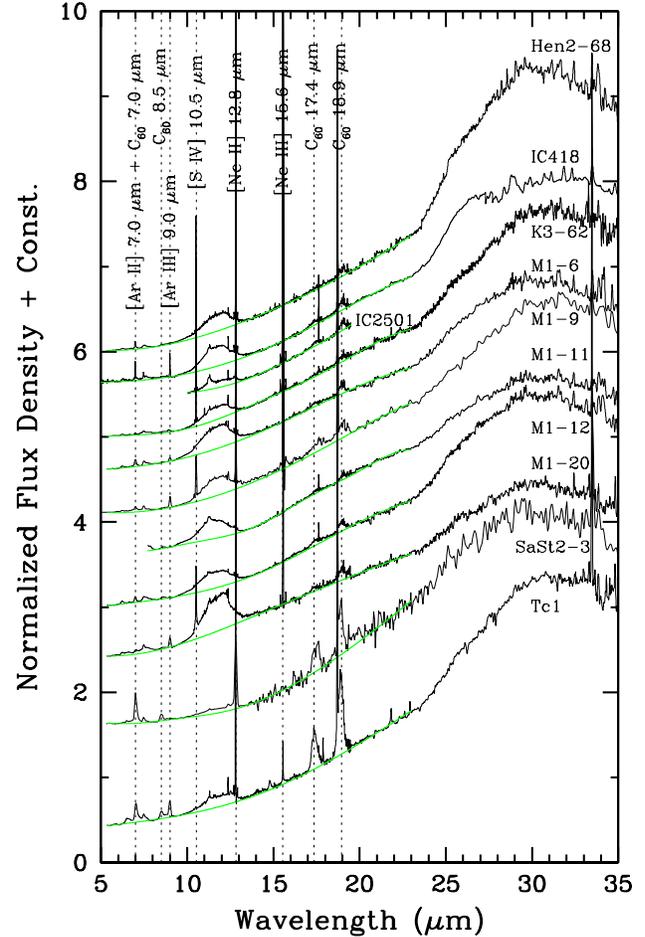}
\caption{\label{Fig:spec-abs}The 5.3-35 $\mu$m \emph{Spitzer}/IRS and
  \emph{ISO} spectra normalized to the flux density at 20 $\mu$m. 
The green lines are base lines determined by fitting the underlying 
feature-less continuum with third to fifth order spline in the range from 5.3 to 23 $\mu$m, except for
 IC2501 and M1-11. For IC2501 we determined the continuum over 2.5-5.0
 $\mu$m and 10-19.6 $\mu$m, while for M1-11 we used 2.5-5.0 and 8-23
 $\mu$m. The flux density normalization for IC2501 was using the
 extrapolated continuum from 10-19.6 $\mu$m continuum (the green line).
Several emission features seen in these spectra are indicated by dotted
 lines. The 17.4 and 18.9 $\mu$m C$_{60}$ bands
 are also presented in Fig. \ref{Fig:spec-16-20um}. The 7.0 and 8.5 $\mu$m
 C$_{60}$ bands and the 6-9 $\mu$m PAH bands are also presented in Fig. \ref{Fig:spec-5-9um}. 
}
\end{figure}

\begin{figure*}
\includegraphics[width=\textwidth,bb=27 307 573 693,clip]{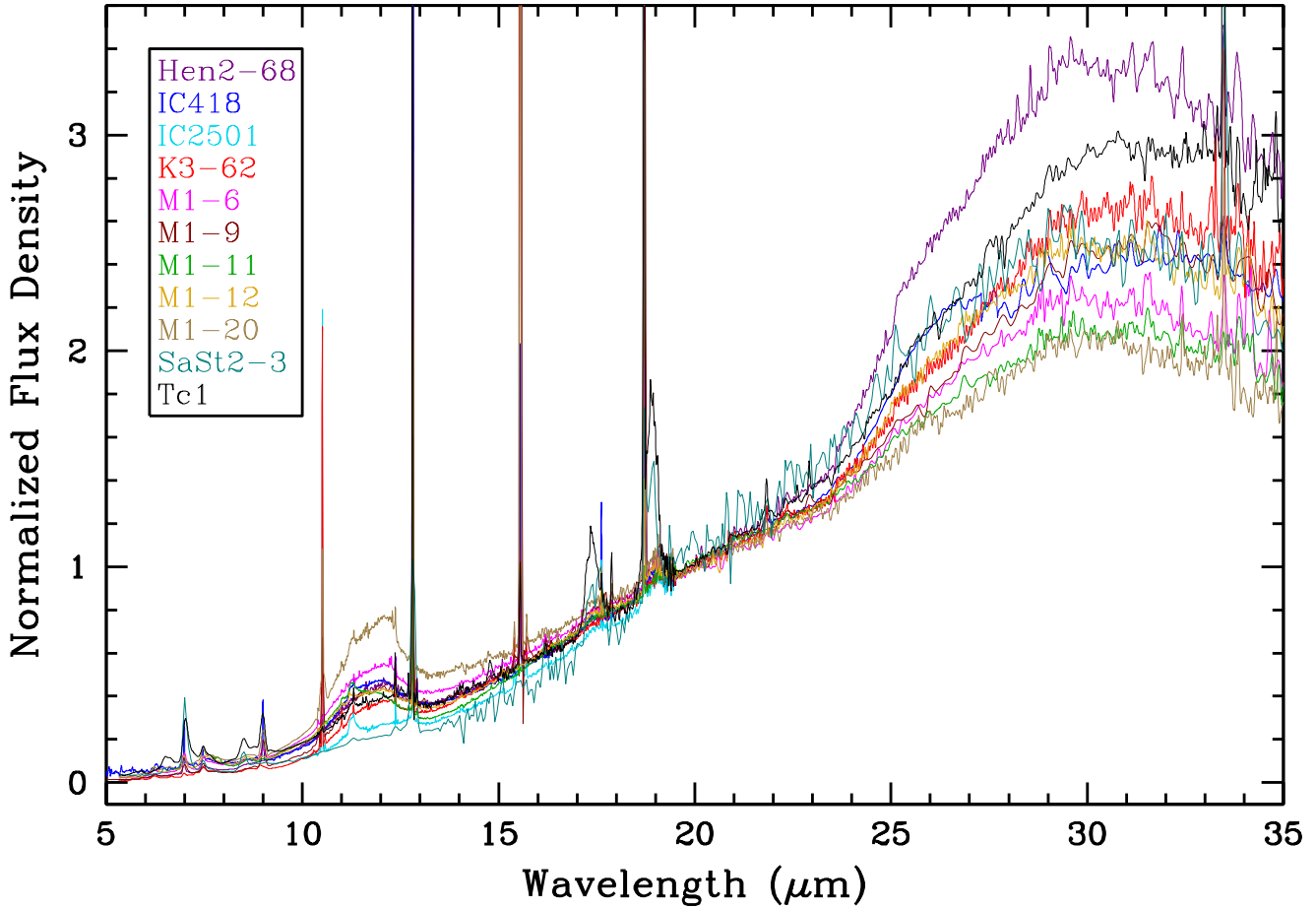}
\caption{\label{Fig:spec-norm20um}The \emph{Spitzer}/IRS and
  \emph{ISO} spectra normalized to the flux density at 20 $\mu$m.}
\end{figure*}

The 5.3-35 $\mu$m spectra are presented in Fig. \ref{Fig:spec-abs}. 
The data of M1-11 shortwards of 10 $\mu$m are from ESO 
VLT/VISIR (See \citealt{Otsuka_2013}). 
The positions of the C$_{60}$ bands and several atomic lines 
seen in the spectra are indicated by the dotted lines. 
The blue edge of the SH spectrum of
SaSt2-3 is very noisy (SNR$\lesssim$5). Therefore, we combined the 5.3-14
$\mu$m SL, $>$13.5 $\mu$m SH, and LH data into a single spectrum. 
In the range between 5.3 and 23 $\mu$m, we fitted the underlying 
feature-less continuum by a third to fifth order single spline function 
indicated by the green lines. Since there is no spectral
data between 5-7 $\mu$m for M1-11 and $<$10 $\mu$m for IC2501, we used the
\emph{AKARI}/IRC 2.5-5.0 $\mu$m spectra as anchor data at the 
shortest wavelength. We subtracted the spline 
component from the data before measuring the fluxes of the C$_{60}$ lines. 
The details of the \emph{AKARI}/IRC 2.5-5.0
$\mu$m spectra are discussed in Section \ref{S:akari}.

There are strong variations in the intensity of the 8.5, 17.4, and 18.9 
$\mu$m C$_{60}$ bands relative to the local dust continuum indicated by
the green lines, with SaSt2-3 and Tc1 showing especially strong 
8.5, 17.4, and 18.9 $\mu$m C$_{60}$ bands.

The 6-9 $\mu$m PAH bands are seen in all PNe, although the band-profiles are
different from ones frequently seen in non-C$_{60}$ PNe. There is the 
possibility that the 6.99 and 7.07 $\mu$m  C$_{70}$ resonances 
might contribute to the 7.0 $\mu$m band. However, we could 
not detect any C$_{70}$ bands in our sample, except for the already
reported features in Tc1 
\citep{Cami_2010}. Therefore, in the PNe in our sample except for Tc1, the 7.0
$\mu$m line is concluded to be due to a complex of the 6.92 $\mu$m aliphatic 
vibration band, the [Ar\,{\sc ii}] 6.99 $\mu$m, and the 7.0 $\mu$m
C$_{60}$ band.
Except for SaSt2-3 and Tc1, the 8.5 $\mu$m C$_{60}$ bands are 
contaminated by the 8.6 $\mu$m PAH band. We focus on the 17.4 and
18.9 $\mu$m C$_{60}$ bands and on the 6-9 $\mu$m complex in Sections
\ref{s_c60_17_18} and \ref{5-9pah}, respectively.

The broad 11 and 30 $\mu$m features are also seen in all PNe. 
Fig. \ref{Fig:spec-norm20um} shows the spectra normalized to the flux
density at 20 $\mu$m. All the spectra show a thermal 
dust continuum; the slope (clearly visible over 12-23 $\mu$m) is 
different for the different objects, pointing to different dust 
temperatures, caused by the difference in irradiation. The strengths of
the 17.4 $\mu$m and 18.9 $\mu$m C$_{60}$ bands are not changing with the
local dust continuum. We investigate correlations between the C$_{60}$ band
strength, the broad 11 $\mu$m band, and the local dust continuum in the 
following sections.

\subsection{The 17.4 and 18.9 $\mu$m C$_{60}$ line profiles \label{s_c60_17_18}}
\begin{figure}
\centering
\resizebox{\hsize}{!}{
\includegraphics[clip,bb=26 212 585 713]{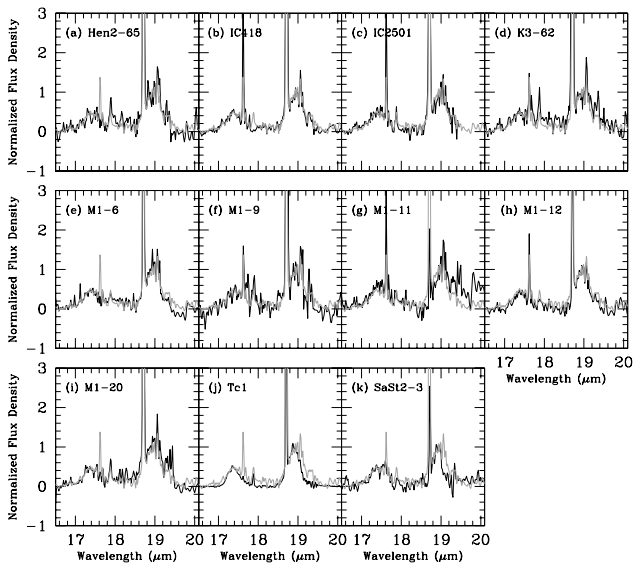}
}
\caption{\label{Fig:spec-16-20um}The local continuum subtracted spectra
 over 16.5-20.1
  $\mu$m, normalized to the peak flux density 
in the 18.9 $\mu$m C$_{60}$ band. The gray lines represent the observed spectra of
 each object and the grey lines show the average spectrum among all
 objects.}
\end{figure}

Fig. \ref{Fig:spec-16-20um} displays the continuum subtracted spectra
over 16.5-20.1 $\mu$m, showing the 17.4 and 18.9 $\mu$m C$_{60}$
features in all objects in our sample. 
Narrow emission lines are also seen in the same range; e.g., 
H~{\sc i} 17.61 $\mu$m ($n$=11-18), [P\,{\sc iii}] 17.89 $\mu$m 
($^{2}P_{1/2}$-$^{2}P_{3/2}$), [S\,{\sc iii}] 18.71 $\mu$m 
($^{3}P_{1}$-$^{3}P_{2}$), and H~{\sc i} 19.06 $\mu$m ($n$=7-8), as
well as spike noise. 
Since the red-wings of the 18.9 $\mu$m C$_{60}$ bands in M1-11 and M1-20 
are noisy, the full width at zero intensity of the 18.9 $\mu$m C$_{60}$
band is a little bit larger than that in the others. The narrower width of 
the 18.9 $\mu$m C$_{60}$ band in SaSt2-3 relative to the other objects is due to the
low SNR around 19 $\mu$m.

Although there are some issues in the data quality as explained above, 
the normalized line profiles of the 17.4 and 18.9 $\mu$m 
C$_{60}$ bands are almost similar to each other, as can be seen from the
average spectrum, which is plotted in grey in each panel.

\subsection{The 7.0 and 8.5 $\mu$m C$_{60}$ and the broad 
6-9 $\mu$m complex \label{5-9pah}} 
\begin{figure}
\centering
\resizebox{\hsize}{!}{
\includegraphics[clip,bb=25 158 408 713]{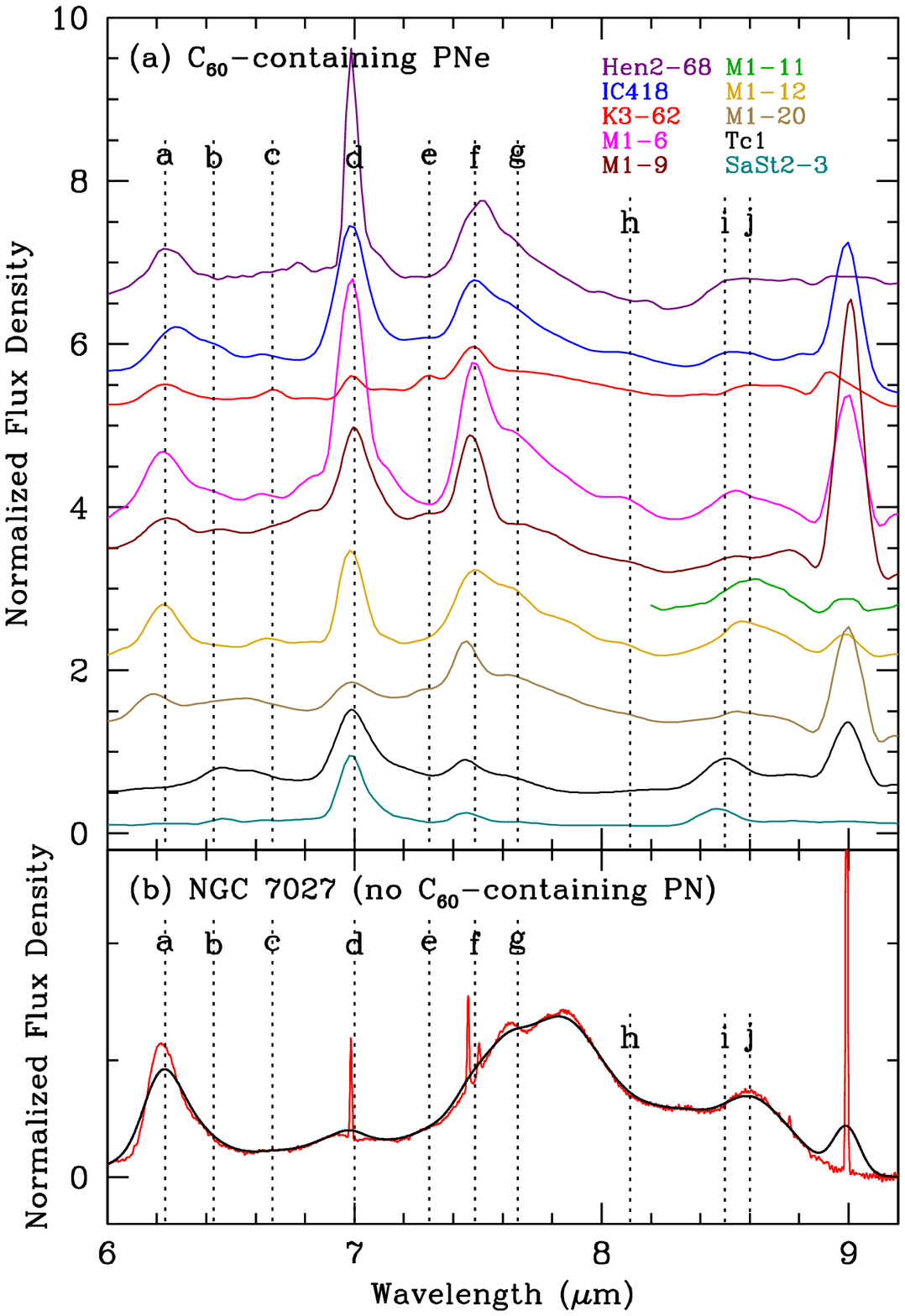}
}
\caption{\label{Fig:spec-5-9um} (upper panel) 
The continuum subtracted over 6.0-9.2 
$\mu$m of all sources for which data is available in this range, 
normalized to the peak flux density of the 
8.5 $\mu$m C$_{60}$+8.6 $\mu$m PAH band and then adjusted. 
The emission features seen in these spectra 
are indicated by dotted lines and lower-case letters.  
(lower panel) The \emph{ISO}/SWS 
spectrum of the C-rich PN NGC7027. We removed the high-excitation lines such as [Cl\,{\sc v}] 6.71
$\mu$m, He\,{\sc ii} 6.95 $\mu$m, [Na\,{\sc iii}] 7.32 $\mu$m, [Ne\,{\sc
vi}] 7.65 $\mu$m, [Ar\,{\sc v}] 7.90 $\mu$m, [Na\,{\sc vi}] 8.61 $\mu$m,
and [K\,{\sc vi}] 8.83 $\mu$m and the pure rotational transition lines of molecular hydrogen
(H$_{2}$) lines from the original spectrum of NGC7027. 
The resolution of 
this spectrum was reduced to that of \emph{Spitzer}/IRS.
The original resolution spectrum is indicated by the red line and the 
degraded one is indicated by the black line. 
The emission features seen in these spectra are indicated by dotted lines 
and lower-case letters. a: PAH C=C stretch 6.23 $\mu$m, b: C$_{60}$
 $^{+}$ 6.4 $\mu$m (possible), 
c: C$_{24}$ planar (possible) or PAH C=C stretch 6.67 $\mu$m, d: C$_{60}$ 7.0 $\mu$m + [Ar\,{\sc ii}] 
6.99 $\mu$m, e: an emission feature at 7.30 $\mu$m seen by \citet{Scott:1996aa,Scott:1997ab}, 
f: C$_{60}^{+}$ + H~{\sc i} 7.46 $\mu$m and the H~{\sc i} 7.50 $\mu$m 
complex, g: PAH C-C stretch 7.7 $\mu$m, h: C$_{60}^{+}$ 8.1 $\mu$m
 (possible), 
i: C$_{60}$ 8.5 $\mu$m, and j: PAH C-H in-plane bend 8.6 $\mu$m.
}
\end{figure}

\begin{table*}
\caption{3,3, 6.2, 8.6, and 11.3 $\mu$m PAH band fluxes. \label{Table:fluxPAH}}
\begin{tabularx}{\textwidth}{@{}lccccccccl@{}}
\hline\hline
{Nebula}&
{$\lambda_{\rm obs}$}&
{$F$(PAH 3.3)}&
{$\lambda_{\rm obs}$}&
{$F$(PAH 6.2)}&
{$\lambda_{\rm obs}$}&
{$F$(PAH 8.6)}&
{$\lambda_{\rm obs}$}&
{$F$(PAH 11.3)}&
{$F$(3.3)/}\\
{}&
{($\mu$m)}&
{(erg s$^{-1}$ cm$^{-2}$)}&
{($\mu$m)}&
{(erg s$^{-1}$ cm$^{-2}$)}&
{($\mu$m)}&
{(erg s$^{-1}$ cm$^{-2}$)}&
{($\mu$m)}&
{(erg s$^{-1}$ cm$^{-2}$)}&
{$F$(11.3)}
\\
\hline
Hen2-68&3.30&1.38(--13)$\pm$4.83(--14)  & 6.23 & 5.74(--13)$\pm$1.04(--14) & 8.70 & 2.09(--13)$\pm$4.68(--15) & 11.26 & 2.02(--13)$\pm$3.39(--14) & 0.68$\pm$0.12\\ 
IC418  &3.29&1.96(--11)$\pm$1.11(--12) & 6.27 & 3.46(--11)$\pm$2.03(--13) & 8.68 & 1.62(--11)$\pm$5.45(--14) & 11.26 & 2.74(--11)$\pm$8.62(--13) & 0.72$\pm$0.23\\ 
IC2501 &3.30&3.18(--12)$\pm$1.16(--13) &  &  &  &  &     11.28 & 5.94(--12)$\pm$3.74(--13) &0.54$\pm$0.04      \\ 
K3-62  &3.30&2.86(--13)$\pm$3.46(--14) & 6.23 & 1.41(--12)$\pm$2.99(--14) & 8.72 & 7.38(--13)$\pm$5.96(--15) & 11.25 & 8.94(--13)$\pm$6.03(--14) & 0.32$\pm$0.04\\ 
M1-6   &3.30&4.05(--13)$\pm$4.02(--14) & 6.23 & 2.58(--12)$\pm$3.33(--14) & 8.66 & 5.50(--13)$\pm$8.25(--14) & 11.27 & 5.81(--13)$\pm$8.75(--14) & 0.70$\pm$0.13\\ 
M1-9   && & 6.23 & 1.75(--13)$\pm$1.56(--15) & 8.71 & 6.28(--14)$\pm$1.30(--14) & 11.28 & 1.25(--13)$\pm$1.87(--14) & \\ 
M1-11  &3.29 &7.02(--12)$\pm$2.89(--13)  &  &  & 8.69 & 5.79(--12)$\pm$2.95(--13) & 11.29 & 1.16(--11)$\pm$7.85(--13) &0.61$\pm$0.04 \\ 
M1-12  &3.29&5.41(--13)$\pm$3.96(--14) & 6.22 & 3.26(--13)$\pm$7.04(--14) & 8.69 & 8.60(--13)$\pm$1.24(--13) & 11.25 & 8.61(--13)$\pm$1.23(--13) & 0.63$\pm$0.10\\ 
M1-20  && & 6.22 & 7.68(--13)$\pm$2.23(--14) & 8.71 & 3.98(--13)$\pm$8.54(--14) & 11.26 & 8.56(--13)$\pm$1.63(--13) & \\ 
SaSt2-3&& & 6.25 & 1.49(--14)$\pm$9.90(--16) &  &    & 11.29 & 1.36(--14)$\pm$4.68(--16) &  \\ 
Tc1    && & 6.20 & 8.08(--13)$\pm$5.42(--14) &  &    & 11.27 &
			 1.40(--12)$\pm$2.20(--13) & \\
\hline
\end{tabularx}
\end{table*}

We investigate the 6-9.2 $\mu$m continuum-subtracted spectra 
(Fig. \ref{Fig:spec-abs}), and the result is shown in the upper panel of Fig. \ref{Fig:spec-5-9um}. 
As a reference, we show the 6-9.2 $\mu$m line profile of the C-rich
young PN NGC7027 in the lower panel of Fig. \ref{Fig:spec-5-9um}. 
We did not detect C$_{60}$ bands in this PN. According to the
classification of 6-9 $\mu$m PAH profiles by \citet{Peeters02}, NGC7027 
is a class B PN. \citet{Bernard-Salas_09_LMC} classified
the 6-9 $\mu$m PAH profiles of 11 Magellanic Cloud PNe with no C$_{60}$
as class B among 14 objects. Hence a comparison between our sources and
NGC7027 is useful to investigate the differences between
C$_{60}$ and non-C$_{60}$ PNe and also to see the contribution 
from the 6.92 $\mu$m aliphatic emission and from the 8.6 $\mu$m PAH emission to the 7.0 $\mu$m and 8.5 
$\mu$m C$_{60}$ bands, respectively.

The line-profiles in the C$_{60}$-containing PNe are similar to each other, except for 
the 6.2 and 8.5 $\mu$m bands. Only two sources, Tc1 and SaSt2-3 clearly show 
the strong 8.5 $\mu$m C$_{60}$ band, while the 6.2 $\mu$m PAH band is very weak.
In other sources these bands are heavily contaminated by PAHs at 8.6 and
6.2 $\mu$m, respectively (see Fig. \ref{Fig:spec-5-9um}a), thus
hampering the flux measurements of the 7.0 and 8.5 $\mu$m C$_{60}$ bands. The 7.2-8.1
$\mu$m band profile in the C$_{60}$-containing PNe are quite different
from that in NGC7027. While NGC7027 and other class B objects exhibit 7.7
$\mu$m emission peaking between 7.8 and 8 $\mu$m, our objects show
little emission in this wavelength range but instead emit the bulk of 
their emission shortwards of the 7.8 $\mu$m complex. Hence,
despite the H~{\sc i} contamination, this suggests that they likely 
exhibit class A profiles. The carrier of the latter is
believed to be more processed than that of the class B profiles 
\citep{Peeters02,Sloan:2005aa,Sloan:2007aa,Boersma:2008aa}, because the
class A and B profiles tend to be observed in reflection nebulae and H\,{\sc
ii} regions and in AGB stars and PNe, respectively.

A 7.30 $\mu$m emission band is seen in K3-62. 
\citet{Scott:1996aa} and \citet{Scott:1997ab}  
found that an emission feature at this wavelength may be due to 
chemically processed hydrogenated amorphous carbons (HAC) grains. 
HAC is a generic name for a mixture of aliphatic and aromatic carbon, 
consisting of 
PAH clusters embedded within a matrix of aliphatically bonded material.
Moreover, a 7.49 $\mu$m emission band 
is detected in all PNe. This emission band is possibly due to the 7.50
$\mu$m C$_{60}^{+}$ resonance. 
On the other hand, the 7.49 $\mu$m feature will have large contamination
from the H~{\sc i} 7.46 $\mu$m ($n$=6-5) and the H~{\sc i} 7.50 $\mu$m ($n$=8-6). In IC418, for
example, our measured respective fluxes of  H~{\sc i} 7.46 $\mu$m and H~{\sc i} 
7.50 $\mu$m are 1.59(--11) and 5.89(--12) erg s$^{-1}$ cm$^{-2}$. 
The observed intensity ratio of H~{\sc i} 7.46 $\mu$m to 7.50 $\mu$m
(2.70) is smaller than the theoretical value (3.77) in the case 
of $T_{\epsilon}$=10$^{4}$ K and $n_{\epsilon}$=10$^{3}$ cm$^{-3}$
derived by \citet{Storey_1995}. Therefore, we conclude that the 7.49
$\mu$m complex is a combination of both H\,{\sc i} and C$_{60}^{+}$. We assume that only part of the 7.49
$\mu$m feature is due to H\,{\sc i}, and the remainder is carried by C$_{60}^{+}$.

The fluxes of the 6.2 and 8.6 $\mu$m PAH bands are listed in Table \ref{Table:fluxPAH}.

\subsection{Flux measurements of the C$_{60}$ lines \label{S:flux-measurement}}

\begin{table*}
\caption{Results of the Gaussian fits for the C$_{60}$ lines.\label{Table:fluxc60}}
\begin{tabularx}{\textwidth}{@{}lrcl@{\hspace{65pt}}lrcl@{}} 
\hline\hline
{Nebula}&
{$\lambda_{obs}$}&
{FWHM}&
{Flux} 
{}&
{Nebula}&
{$\lambda_{obs}$}&
{FWHM}&
{Flux} \\
{}&
{($\mu$m)}&
{($\mu$m)}&
{(erg s$^{-1}$ cm$^{-2}$)}&
{}&
{($\mu$m)}&
{($\mu$m)}&
{(erg s$^{-1}$ cm$^{-2}$)}\\
\hline
Hen2-68	&	6.98	&	5.99(--2)	$\pm$	1.54(--3)	&	1.06(--12)	$\pm$	4.01(--14)$^{a}$	&	M1-11	&	8.55	&	1.89(--1)	$\pm$	9.51(--3)	&	3.82(--12)	$\pm$	3.46(--13)$^{c}$	\\
	&	 	&				&	 9.78(--13)	$\pm$	4.28(--14)$^{b}$	&		&		&				&	9.61(--12)	$\pm$	4.55(--13)$^{d}$	\\
	&	8.51	&	1.87(--1)	$\pm$	5.66(--3)	&	2.16(--13)	$\pm$	7.72(--15)$^{c}$	&		&	17.43	&	3.56(--1)	$\pm$	6.52(--2)	&	3.60(--12)	$\pm$	7.74(--13)	\\
	&	 	&				&	4.25(--13)	$\pm$	9.02(--15)$^{d}$	&		&	19.02	&	3.96(--1)	$\pm$	5.14(--2)	&	6.89(--12)	$\pm$	1.16(--12)	\\
	&	17.43	&	4.17(--1)	$\pm$	1.12(--2)	&	3.95(--13)	$\pm$	5.03(--14)	&	M1-12	&	6.99	&	1.13(--1)	$\pm$	1.80(--3)	&	3.69(--12)	$\pm$	7.27(--14)$^{a}$	\\
	&	18.96	&	4.49(--1)	$\pm$	4.90(--2)	&	7.83(--13)	$\pm$	1.34(--13)	&		&		&				&	2.35(--12)	$\pm$	1.06(--14)$^{b}$	\\
IC418	&	7.01	&	1.95(--1)	$\pm$	1.01(--2)	&	8.62(--11)	$\pm$	2.74(--12)$^{a}$	&		&	8.55	&	2.18(--1)	$\pm$	1.03(--2)	&	1.24(--12)	$\pm$	9.58(--14)$^{c}$	\\
	&	 	&				&	 3.28(--11)	$\pm$	3.07(--12)$^{b}$	&		&		&				&	2.10(--12)	$\pm$	1.02(--13)$^{d}$	\\
	&	8.51	&	1.45(--1)	$\pm$	1.37(--2)	&	 5.75(--12)	$\pm$	6.67(--13)$^{c}$	&		&	17.40	&	3.59(--1)	$\pm$	2.14(--2)	&	1.09(--12)	$\pm$	1.09(--13)	\\
	&	 	&				&	 2.20(--11)	$\pm$	8.61(--13)$^{d}$	&		&	18.94	&	4.02(--1)	$\pm$	1.63(--2)	&	2.41(--12)	$\pm$	1.20(--13)	\\
	&	17.38	&	3.48(--1)	$\pm$	1.12(--2)	&	2.04(--11)	$\pm$	7.93(--13)	&	M1-20	&	7.02	&	1.57(--1)	$\pm$	2.60(--3)	&	7.91(--13)	$\pm$	2.37(--14)$^{a}$	\\
	&	18.93	&	4.34(--1)	$\pm$	1.86(--2)	&	4.44(--11)	$\pm$	2.04(--12)	&		&		&				&	6.83(--13)	$\pm$	1.32(--14)$^{b}$	\\
IC2501	&	17.44	&	3.49(--1)	$\pm$	2.53(--2)	&	1.97(--12)	$\pm$	1.93(--13)	&		&	8.53	&	2.29(--1)	$\pm$	3.73(--2)	&	3.67(--13)	$\pm$	8.21(--14)$^{c}$	\\
	&	19.02	&	3.78(--1)	$\pm$	4.82(--2)	&	3.93(--12)	$\pm$	6.08(--13)	&		&		&				&	6.45(--13)	$\pm$	1.43(--13)$^{d}$	\\
K3-62	&	6.99	&	7.94(--2)	$\pm$	1.33(--3)	&	6.05(--13)	$\pm$	1.77(--14)$^{a}$	&		&	17.38	&	3.71(--1)	$\pm$	3.88(--2)	&	4.36(--13)	$\pm$	5.39(--14)	\\
 	&	 	&				&	4.20(--13)	$\pm$	3.74(--14)$^{b}$	&		&	18.98	&	3.58(--1)	$\pm$	2.80(--2)	&	8.74(--13)	$\pm$	9.53(--14)	\\
	&	8.56	&	1.26(--1)	$\pm$	4.43(--3)	&	2.46(--13)	$\pm$	1.12(--14)$^{c}$	&	SaSt2-3	&	7.01	&	1.23(--1)	$\pm$	2.16(--3)	&	4.52(--13)	$\pm$	1.28(--14)$^{a}$	\\
	&	 	&				&	9.83(--13)	$\pm$	1.27(--14)$^{d}$	&		&		&				&	3.15(--13)	$\pm$	1.98(--14)$^{b}$	\\
	&	17.35	&	3.25(--1)	$\pm$	6.29(--2)	&	6.04(--13)	$\pm$	1.25(--14)	&		&	8.50	&	1.71(--1)	$\pm$	9.00(--4)	&	1.19(--13)	$\pm$	7.77(--16)	\\
	&	18.99	&	4.43(--1)	$\pm$	4.79(--2)	&	1.32(--12)	$\pm$	1.66(--13)	&		&	17.39	&	3.11(--1)	$\pm$	5.80(--2)	&	1.68(--13)	$\pm$	3.55(--14)	\\
M1-6	&	6.99	&	1.34(--1)	$\pm$	2.04(--3)	&	5.50(--12)	$\pm$	1.40(--14)$^{a}$	&		&		&				&	1.30(--13)	$\pm$	2.75(--14)$^{e}$	\\
	&		&				&	4.88(--12)	$\pm$	1.51(--13)$^{b}$	&		&	18.92	&	2.92(--1)	$\pm$	1.77(--2)	&	3.03(--13)	$\pm$	2.40(--14)	\\
	&	8.51	&	1.74(--1)	$\pm$	2.08(--3)	&	5.09(--13)	$\pm$	9.18(--14)$^{c}$	&		&	 	&				&	2.74(--13)	$\pm$	2.17(--14)$^{e}$	\\
	&		&				&	1.06(--12)	$\pm$	1.24(--13)$^{d}$	&	Tc1	&	7.03	&	1.66(--1)	$\pm$	4.44(--3)	&	1.75(--11)	$\pm$	5.70(--13)$^{a}$	\\
	&	17.36	&	3.91(--1)	$\pm$	5.67(--2)	&	9.45(--13)	$\pm$	1.66(--13)	&		&		&				&	1.67(--11)	$\pm$	5.86(--13)$^{b}$	\\
	&	18.90	&	3.56(--1)	$\pm$	1.62(--2)	&	1.98(--12)	$\pm$	1.07(--13)	&		&	8.50	&	1.90(--1)	$\pm$	3.00(--3)	&	6.22(--12)	$\pm$	1.34(--14)	\\
M1-9	&	7.00	&	1.36(--1)	$\pm$	2.95(--3)	&	3.47(--13)	$\pm$	8.16(--15)$^{a}$	&		&	17.38	&	3.66(--1)	$\pm$	1.08(--2)	&	1.38(--11)	$\pm$	5.00(--13)	\\
	&		&				&	2.94(--13)	$\pm$	8.62(--15)$^{b}$	&		&	 	&				&	1.07(--11)	$\pm$	3.87(--13)$^{e}$	\\
	&	8.54	&	2.03(--1)	$\pm$	8.70(--3)	&	5.93(--14)	$\pm$	2.79(--15)$^{c}$	&		&	18.90	&	3.47(--1)	$\pm$	4.39(--3)	&	2.40(--11)	$\pm$	4.42(--13)	\\
	&		&				&	1.22(--13)	$\pm$	1.33(--14)$^{d}$	&		&	 	&				&	2.17(--11)	$\pm$	3.99(--13)$^{e}$	\\
	&	17.40	&	3.53(--1)	$\pm$	5.23(--2)	&	2.15(--13)	$\pm$	3.77(--14)	&		&		&				&				\\
	&	18.89	&	3.73(--1)	$\pm$	2.19(--2)	&	4.19(--13)	$\pm$	3.32(--14)	&		&		&				&				\\
\hline
\end{tabularx}
\raggedright
\begin{tabular}{l}
{$^{a}$}{Total flux of the 7.0 $\mu$m line.}\\
{$^{b}$}{The residual flux after the predicted [Ar\,{\sc ii}]
 6.99 $\mu$m flux and 6.92 $\mu$m aliphatic band feature are subtracted. The value is the flux
 of the C$_{60}$ 7.0 $\mu$m feature.}\\
{$^{c}$}{The flux of C$_{60}$ 8.5 $\mu$m feature measured using
 a two-component Gaussian fit. For SaSt2-3 and Tc1, a single Gaussian is fitted to
 the  C$_{60}$ 8.5 $\mu$m feature.}\\
{$^{d}$}{Total flux of the C$_{60}$ 8.5 $\mu$m and PAH 8.6
 $\mu$m bands.}\\
{$^{e}$}{by applying the flux ratio of C$_{60}$/C$_{70}$ observed
 in Tc1 by \citet{Cami_2010}.}
\end{tabular}
\end{table*}

\begin{table}
\caption{The flux ratios of the C$_{60}$ lines relative to the 18.9
 $\mu$m C$_{60}$ band. \label{Table:ratio}}
\begin{tabularx}{\columnwidth}{@{}XlYcYcYc@{}}
\hline\hline
{Nebula} & 
{$F$(7.0)/$F$(18.9)}  & 
{$F$(8.5)/$F$(18.9)}  & 
{$F$(17.4)/$F$(18.9)} \\
\hline
Hen2-68	&	1.25 $\pm$	0.22	&	0.28 $\pm$	0.05
 &	0.50 $\pm$	0.11	\\
IC418	&	0.74 $\pm$	0.08	&	0.50 $\pm$	0.03
 &	0.46 $\pm$	0.03	\\
IC2501	&	 	&	 	&	0.50 $\pm$	0.05	\\
K3-62	&	0.32 $\pm$	0.05	&	0.19 $\pm$	0.02
 &	0.46 $\pm$	0.06	\\
M1-6	&	2.46 $\pm$	0.15	&	0.26 $\pm$	0.05
 &	0.48 $\pm$	0.09	\\
M1-9&	0.70 $\pm$	0.06	&	0.14 $\pm$	0.01
 &	0.51 $\pm$	0.10	\\
M1-11	&	 	 	&	0.55 $\pm$	0.11
 &	0.52 $\pm$	0.14	\\
M1-12	&	0.98 $\pm$	0.05	&	0.51 $\pm$	0.05
 &	0.45 $\pm$	0.05	\\
M1-20	&	0.78 $\pm$	0.09	&	0.42 $\pm$	0.10
 &	0.50 $\pm$	0.08	\\
SaSt2-3	&	1.04 $\pm$	0.11	&	0.39 $\pm$	0.07
 &	0.55 $\pm$	0.13	\\
Tc1	&	0.77 $\pm$	0.03	&	0.29 $\pm$	0.01
 &	0.49 $\pm$	0.02\\	
\hline
\end{tabularx}
\end{table}

We employed multiple Gaussian fitting for all C$_{60}$ bands to separate
the various components in each feature. The
details are discussed in the Appendix and the results are listed in Table \ref{Table:fluxc60}. We also
list the total flux of the 7.0 $\mu$m complex composed of the 7.0 $\mu$m
C$_{60}$ band, the 6.92 $\mu$m aliphatic band, and the [Ar\,{\sc ii}] 6.99
$\mu$m line, as well as the total flux of the 8.5 $\mu$m C$_{60}$ and 8.6 $\mu$m PAH complex. 
The total fluxes can be treated as upper limits to the C$_{60}$
fluxes. In the fits of the 17.4 $\mu$m and 18.9 $\mu$m C$_{60}$ bands, we employed two or 
three component Gaussians.

The measurement errors of the central wavelength, the FWHM, and the fluxes 
of the C$_{60}$ lines are 1-$\sigma$. The 17.4 $\mu$m and 18.9 $\mu$m
C$_{60}$ bands in Tc1 include a contribution from C$_{70}$
\citep{Cami_2010}, with the flux ratios of the
C$_{60}$ to C$_{70}$ at 17.4 and 18.9 $\mu$m of 3.44 and 9.29,
respectively. For Tc1, we also listed the fluxes of the 17.4 and 18.9
$\mu$m C$_{60}$ lines after the contribution from C$_{70}$ lines are
subtracted. For SaSt2-3, the fluxes of 17.4 and 18.9 $\mu$m C$_{60}$ are
also listed when the above ratios for Tc1 are applied, although we could
not clearly detect any C$_{70}$ bands in SaSt2-3.

The flux ratios relative to 18.9 $\mu$m C$_{60}$ band are summarized in 
Table \ref{Table:ratio}. In Tc1, the 7.0 $\mu$m feature is a complex of
the C$_{60}$ and C$_{70}$, but the exact fraction of each component to the 
7.0 $\mu$m band is unknown. Therefore, the $F$(7.0 $\mu$m)/$F$(18.9
$\mu$m) value in Tc 1 represents an upper limit. In other PNe, since we could not 
detect any C$_{70}$ emissions, the contribution of the 
C$_{70}$ to the 7.0 $\mu$m band is very small. 
The $F$(17.4 $\mu$m)/$F$(18.9 $\mu$m) ratio is almost constantly
$\sim$0.5 in all objects in our sample as we argued in Section
\ref{s_c60_17_18}, while there is more scatter in the $F$(7.0 $\mu$m)/$F$(18.9 $\mu$m) and 
the $F$(8.5 $\mu$m)/$F$(18.9 $\mu$m) ratios due to the contributions
from the [Ar\,{\sc ii}] 6.99 $\mu$m line and the 8.6 $\mu$m PAH band to the 7.0
and 8.5 $\mu$m C$_{60}$ bands, respectively.

\subsection{The 3.3 $\mu$m PAH band in \emph{AKARI} spectra \label{S:akari}}

\begin{figure*}
\centering
\includegraphics[bb=23 370 585 713,width=\textwidth]{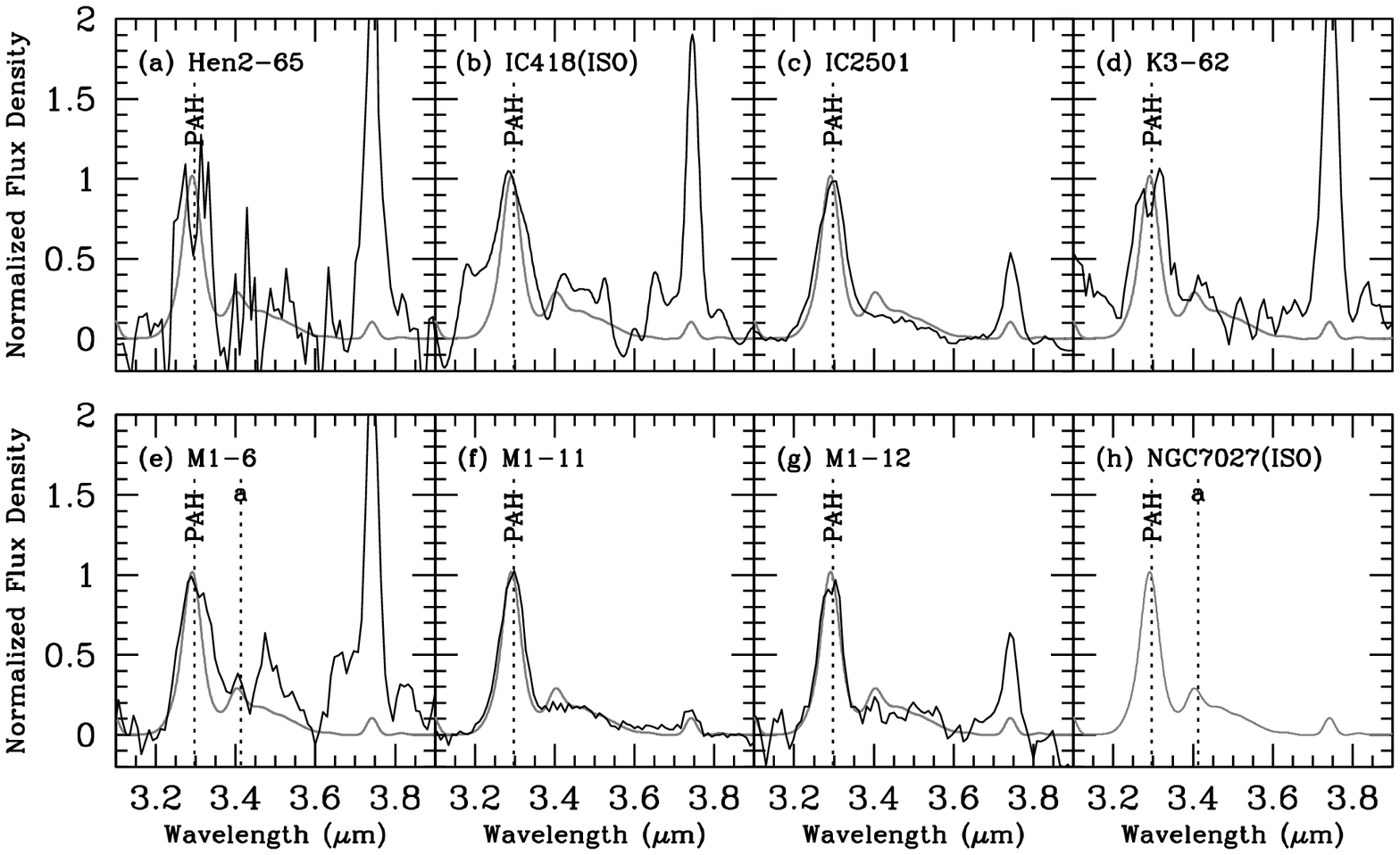}
\caption{\label{akari} 
The \emph{AKARI}/IRC and \emph{ISO} 3.1-3.9 $\mu$m spectra. 
We removed the H~{\sc i} 3.2 $\mu$m line. The grey lines are the
 \emph{ISO}/SWS spectrum of the C-rich PN NGC7027. 
The 3.4-3.6 $\mu$m faint broad band is due to aliphatic hydrocarbon. 
The emission around 3.7 $\mu$m is the H\,{\sc 
i} 3.73 $\mu$m ($n$=5-8). The dashed lines marked with ``a'' in M1-6 and
 NGC7027 represent the 3.38/3.40 $\mu$m aliphatic asymmetric CH$_{3}$, CH$_{2}$ stretch.
} 
\end{figure*}

The reduced \emph{AKARI} spectra show emission bands at
3.2-3.6 $\mu$m, which are due to aromatic and aliphatic hydrocarbon
species. The strong resonance at 3.3 $\mu$m is attributed to the
aromatic C-H stretching vibration 
in PAHs and the 3.4-3.6 $\mu$m faint broad band is due to aliphatic hydrocarbon.
We should note that the 3.3 $\mu$m PAH band is always contaminated with
the H\,{\sc i} 3.2 $\mu$m line ($n$=5-9). 

Using the theoretical intensity ratio
of H~{\sc i} 3.2 $\mu$m to 4.65 $\mu$m ($n$=5-7) = 0.46 in the case of
the electron temperature $T_{\epsilon}$=10$^{4}$ K and the electron
density $n_{\epsilon}$=10$^{4}$ cm$^{-3}$ (\citealt{Storey_1995}), 
we removed the contribution of the H~{\sc i} 3.2 $\mu$m line from 
the 3.3 $\mu$m emission band, assuming that the full width at half
maximum (FWHM) velocities of H~{\sc i} 3.2 and 4.65 $\mu$m are 
the same and the line-profile of these H~{\sc i} lines can be 
represented by a single Gaussian. The resulting spectra in the range from 3.1 to 3.9 $\mu$m are presented in
Fig. \ref{akari}. In panels a-h, as a comparison, we also present the \emph{ISO}/SWS
spectrum of NGC7027 indicated by grey lines. The spectral
resolution of NGC7027 data was reduced down to$\sim$110 at 3.3 $\mu$m using Gaussian convolution 
to match that of \emph{AKARI}/IRC. For the purpose of removing the H~{\sc i} 3.2 $\mu$m, we
used the H~{\sc i} 4.65 $\mu$m in stead of H~{\sc i} 3.73 $\mu$m,
because the SNR of local continuum around 4.65 $\mu$m
is better (IC418, in particular) and there is less contribution from nearby emission-lines.
The positions of some aliphatic hydrocarbons seen in M1-6 and NGC7027 
are indicated by the dotted lines (a). 
The flux density is normalized to the peak intensity
of the 3.3 $\mu$m PAH band and the local continuum of these spectra is
subtracted. Due to low SNR of Hen2-68 and K3-62, we overestimated
the flux of H~{\sc i} 4.65 $\mu$m in both PNe, and we overcorrected the
component from H~{\sc i} 3.2 $\mu$m. The central wavelength and 
the fluxes of the solo 3.3 $\mu$m PAH band error are listed in
the second and third columns of Table \ref{Table:fluxPAH}.

For IC418, we confirmed the presence of the broad PAH 3.3 $\mu$m
emission band in the \emph{ISO} spectrum. The spectral resolution was
also reduced to 110 at 3.3 $\mu$m using Gaussian convolution 
to match that of \emph{AKARI}/IRC.
We again removed the contribution of the H~{\sc i} 3.2 $\mu$m line by
employing the same technique as described above and we again measured the solo
flux of the 3.3 $\mu$m PAH band. The resultant spectrum and the result of the flux
measurement are also shown in Fig. \ref{akari} and Table 
\ref{Table:fluxPAH}, respectively.

The 3.2-3.6 $\mu$m band profile is very similar amongst the sources in
our sample. All our sources exhibit the 3.3
$\mu$m PAH band.

\subsection{The size of PAHs} 


\begin{table}
\caption{The 3.3 and 11.3 $\mu$m PAH band fluxes in \emph{ISO}/SWS
 spectra of no C$_{60}$-containing PNe.}
\begin{tabularx}{\columnwidth}{@{}lccc@{}}
\hline\hline
Nebula &$F$(PAH 3.3)&$F$(PAH 11.3)&$F$(3.3)/$F$(11.3)\\
       &(erg s$^{-1}$ cm$^{-2}$)&(erg s$^{-1}$ cm$^{-2}$)\\
\hline
BD+30$^{\circ}$ 3639 & 6.28(--14) & 2.34(--13) & 0.27 \\ 
Hubble5 & 1.88(--14) & 9.02(--14) & 0.21 \\ 
Hen2-113 & 5.62(--14) & 1.51(--13) & 0.37 \\ 
Hen3-1333 & 8.77(--14) & 2.82(--13) & 0.31 \\ 
M1-78 & 2.10(--14) & 8.32(--14) & 0.25 \\ 
M2-43 & 1.92(--14) & 4.66(--14) & 0.41 \\ 
NGC6302 & 8.67(--15) & 3.36(--14) & 0.26 \\ 
NGC6537 & 4.63(--15) & 1.60(--14) & 0.29 \\ 
NGC7027 & 2.05(--13) & 1.58(--12) & 0.13 \\ 
Vo1 & 1.41(--14) & 3.95(--14) & 0.36 \\ 
\hline
\end{tabularx}
\label{iso}
\end{table}

The 3.3/11.2 $\mu$m PAH band ratio depends on the hardness of the
radiation field (i.e. the average photon energy that is absorbed) 
and the PAH size distribution since 1) the intrinsic PAH flux ratio 
$F$(3.3 $\mu$m)/$F$(11.3 $\mu$m) decreases with increasing number 
of the C-atoms and 2) the 3.3 $\mu$m PAH arises mainly from PAHs 
containing between roughly 30 and 70 C atoms while the 11.2 $\mu$m 
PAH is dominated by PAHs containing between about 80 to several 
hundred carbon atoms (\citealt{Ricca:2012aa,Draine:2007aa,
Schutte:1993aa,Allamandola:1989aa}). The observed $F$(3.3 $\mu$m)/
$F$(11.3 $\mu$m) ratios are given in the last column of 
Table \ref{Table:fluxPAH}. The average $F$(3.3 $\mu$m)/$F$(11.3 $\mu$m) 
is 0.60 among our sample. We also measured the $F$(3.3 $\mu$m)/$F$(11.3
$\mu$m) PAH intensity ratios of 10 other Galactic non-C$_{60}$ 
PNe using their \emph{ISO}/SWS spectra. These results are listed in
Table \ref{iso}. The average value is 0.29, consistent with the 
results by R.~Ohsawa based on \emph{AKARI}/IRC and \emph{Spitzer}/IRS 
spectra of 16 Galactic non-C$_{60}$ PNe (0.3; in private
communication). Since the radiation field is probably roughly similar 
for all C$_{60}$-containing PNe and non-C$_{60}$ PNe, the $F$(3.3
$\mu$m)/$F$(11.3 $\mu$m) ratio is likely governed by the PAH size 
distribution. Thus, we conclude that our C$_{60}$-containing PNe 
have relatively smaller PAHs than non-C$_{60}$ PNe.

\subsection{The broad 11 $\mu$m feature}

\begin{figure}
\centering
\includegraphics[bb=20 157 416 718,clip,width=\columnwidth]{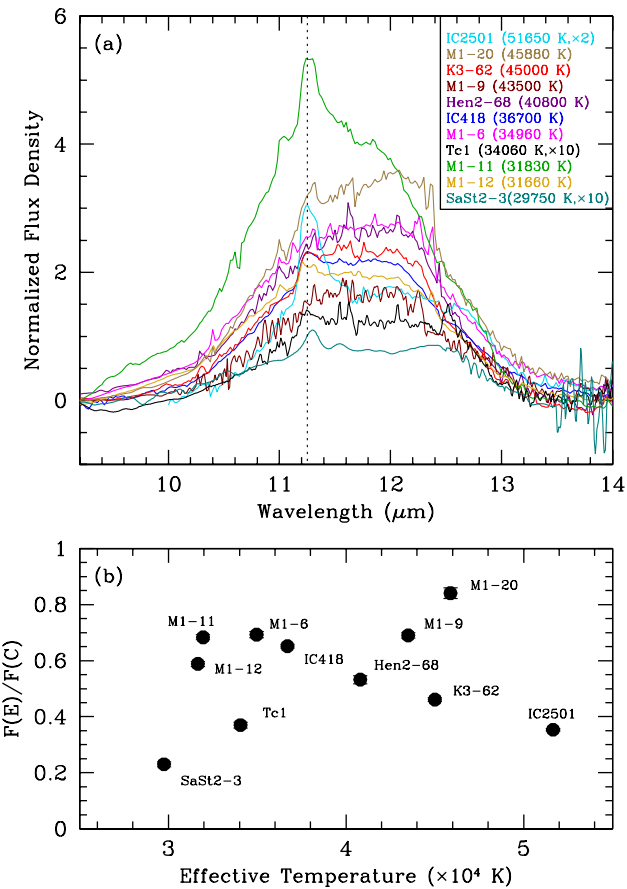}
\caption{\label{Fig:spec-11um} 
({\it upper}) The continuum subtracted spectra over 9.1-14 
$\mu$m, normalized to the peak flux density of C$_{60}$
 18.9 $\mu$m, except for IC2501, SaSt2-3, and Tc1. The scaling factor
 for these PNe are 2, 10, and 10 after the normalization to C$_{60}$
 18.9 $\mu$m. The position of the 11.3 $\mu$m PAH (or HAC) 
is indicated by the dotted line. 
({\it lower}) The plot between the intensity and the effective 
temperatures of the central stars. }
\end{figure}

\begin{figure}
\centering
\resizebox{\hsize}{!}{
\includegraphics[bb=18 155 410 415,clip]{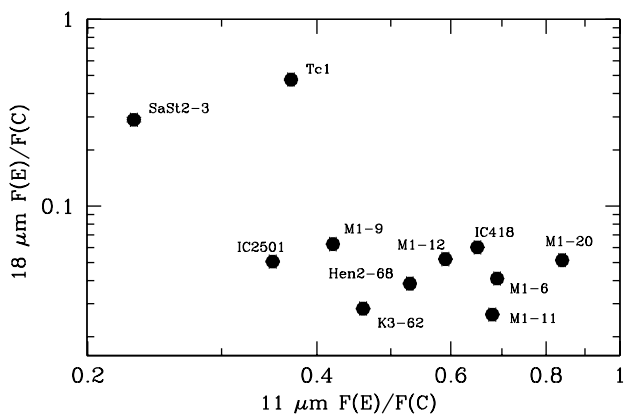}
}
\caption{\label{Fig:spec-18um} 
The correlation between the band strengths of the 18.9 $\mu$m C$_{60}$
 band and the broad 11 $\mu$m band.}
\end{figure}

\begin{table}
\caption{The 11 $\mu$m band fluxes of the emission and continuum components. 
\label{Table:flux11um}}
\begin{tabularx}{\columnwidth}{@{}XlYcYcYc@{}}
\hline\hline
{Nebula}&
{$F$(E)$^{a}$}&
{$F$(C)$^{b}$}&
{$F$(E)/$F$(C)}\\
{}&
{(erg s$^{-1}$ cm$^{-2}$) }&
{(erg s$^{-1}$ cm$^{-2}$) }&
{}\\
\hline
Hen2-68 &2.89(--11) &5.44(--11) &0.53 $\pm$ 0.01\\
IC418   &1.24(--9)    &1.90(--9) &0.65 $\pm$ 0.01\\
IC2501  &5.63(--11)  &1.59(--10) &0.35 $\pm$ 0.01\\
K3-62   &4.91(--11)  &1.06(--10) &0.46 $\pm$ 0.01\\
M1-6    &9.42(--11)   &1.36(--10) &0.69 $\pm$ 0.01\\
M1-9    &9.04(--12)  &2.15(--11) &0.42 $\pm$ 0.01\\
M1-11   &4.22(--10)  &6.17(--10) &0.68 $\pm$ 0.01\\
M1-12   &6.93(--11)  &1.18(--10) &0.59 $\pm$ 0.01\\
M1-20   &4.74(--11)  &5.64(--11) &0.84 $\pm$ 0.02\\
SaSt2-3 &6.20(--13)  &2.54(--12) &0.24 $\pm$ 0.01\\
Tc1     &5.28(--11)    &1.43(--10) &0.37 $\pm$ 0.01\\
\hline
\end{tabularx}
\raggedright
$^{a}${The sum of the band emission without continuum between 10.2 and 13.2 $\mu$m.}\\
$^{b}${The sum of the continuum between 10.2 and 13.2 $\mu$m.}
\end{table}

\begin{table}
\caption{The 18.5-19.3 $\mu$m continuum fluxes and the 18.9 microns C$_{60}$ band strength. 
\label{Table:flux18um}}
\begin{tabularx}{\columnwidth}{@{}XlYcYcYc@{}} 
\hline\hline
{Nebula}&
{$F$(E)$^{a}$}&
{$F$(C)$^{b}$}&
{$F$(E)/$F$(C)}\\
{}&
{(erg s$^{-1}$ cm$^{-2}$) }&
{(erg s$^{-1}$ cm$^{-2}$) }&
{}\\
\hline
Hen2-68 &7.83(--13)  &2.03(--11) &3.85(--2) $\pm$ 6.62(--3)\\
IC418   &4.44(--11)  &7.38(--10) &6.02(--2) $\pm$ 2.82(--3)\\
IC2501  &3.93(--12)  &7.78(--11) &5.05(--2) $\pm$ 7.85(--3)\\
K3-62   &1.32(--12)  &4.66(--11) &2.83(--2) $\pm$ 3.58(--3)\\
M1-6    &1.98(--12)  &4.83(--11) &4.10(--2) $\pm$ 2.28(--3) \\
M1-9    &4.19(--13)  &2.07(--12) &6.26(--2) $\pm$ 5.33(--3) \\
M1-11   &6.89(--12)  &2.62(--11) &2.63(--2) $\pm$ 4.45(--3)\\
M1-12   &2.41(--12)  &4.62(--11) &5.21(--2) $\pm$ 2.71(--3) \\
M1-20   &8.74(--13)  &1.70(--11) &5.14(--2) $\pm$ 5.65(--3)\\
SaSt2-3 &3.03(--13)  &1.04(--12) &2.90(--1) $\pm$ 2.87(--2)\\
Tc1     &2.40(--11)  &5.05(--11) &4.76(--1) $\pm$ 1.01(--2)\\
\hline
\end{tabularx}
\raggedright
$^{a}${The 18.9 $\mu$m C$_{60}$ fluxes.}
$^{b}${The continuum fluxes between 18.5 and 19.3 $\mu$m.}
\end{table}

The appearance of the broad 11 $\mu$m feature is presented in 
Fig. \ref{Fig:spec-11um}a. The flux density is normalized to 
the peak flux density of the 18.9 $\mu$m C$_{60}$ after the continuum
was subtracted. The H~{\sc i} 11.3 
and 12.4 $\mu$m, [S\,{\sc iv}] 10.51 $\mu$m, and [Ne\,{\sc ii}] 
12.81 $\mu$m lines are subtracted out using Multiple Gaussian fitting. 
Except for M1-11, the emission profiles show a nearly flat portion 
between 11.2 and 12.4 $\mu$m. 
The position of the 11.3 $\mu$m PAH band is indicated by the dotted line in
Fig. \ref{Fig:spec-11um}a. The peak intensity of the 11 $\mu$m feature 
in SaSt2-3 and Tc1, which show the prominent C$_{60}$ bands, is much 
weaker than in the other PNe.

The band profile and the peak intensity in IC418, K3-62, M1-6, and M1-12
are very similar. The effective temperatures $T_{\rm eff}$ in these PNe
are quite different (31\,660-45\,000 K), so we believe the 11 $\mu$m
emission does not correlate with $T_{\rm eff}$ (Fig. \ref{Fig:spec-11um}b). 
We measured the integrated flux emission band $F$(E) and the integrated
flux $F$(C) of the underlying continuum between 10.2
and 13.2 $\mu$m (Table \ref{Table:flux11um}). 
The flux of the emission band $F$(E) includes the 11.3 $\mu$m
emission. The error estimate of $F$(E) and $F$(C) is assumed to be equal
to the standard deviation of continuum $<$10.2 and $>$13.2
$\mu$m. Fig. \ref{Fig:spec-11um}b shows the band strengths (defined as
the ratio $F$(E)/$F$(C)) against $T_{\rm eff}$. The correlation
coefficient between these two parameters is 0.10, meaning they are not
correlated. Even if we exclude SaSt2-3 because the SNR of this
object is much lower than those of the other PNe, the correlation 
coefficient is still small (--0.20). Thus, we confirm that there is 
no correlation between the 11 $\mu$m band strength and the $T_{\rm eff}$
of the central star.

The peak intensities of the broad 11 $\mu$m feature in SaSt2-3 and Tc1 are
much lower than those in the other PNe. The relation between the 11 $\mu$m
band and the 18.9 $\mu$m C$_{60}$ band strength is of interest 
(Fig. \ref{Fig:spec-18um}). \citet{Jero:C60excitation} 
demonstrated the similarity between the observed 6-14 $\mu$m spectra 
of C$_{60}$-containing PNe and the spectra of HAC/a-C:H particles, which
shows strong emission around 11.3 $\mu$m (See Fig. 7 of
\citealt{Jero:C60excitation}). They concluded it likely to demonstrate a link between 
the fullerene formation process and the 6-9 and 10-13 $\mu$m broad bands. 
To check the relation 
between the 18 $\mu$m C$_{60}$ and the 11 $\mu$m band strengths, we also measured the
the integrated flux $F$(C) of the underlying continuum between 18.5 and 19.2 $\mu$m, as listed in the third column of Table
\ref{Table:flux18um}. We define the ratio of the 18.9 $\mu$m C$_{60}$
band integrated flux to $F$(C) as the 18.9 $\mu$m C$_{60}$ band strength, and
we plotted the diagram of the 18.9 $\mu$m C$_{60}$ and 11 $\mu$m band
strengths in Fig. \ref{Fig:spec-18um}, although careful treatment of the
18.9 $\mu$m C$_{60}$ band strength of SaSt2-3 is necessary, as we
explained in Section \ref{s_c60_17_18}. No trend exists between 
the 11 $\mu$m band and the 18.9 $\mu$m C$_{60}$ band strengths,
suggesting that the heating mechanism
of the C$_{60}$ and the broad 11 $\mu$m bands is different.

\subsection{The 30 $\mu$m feature and the cold dust continuum}
\label{sec:30mic}

All the sources in our sample show a clear broad emission feature
around 30 $\mu$m, except for IC2501, for which we do not have any data 
$>$20 $\mu$m. The feature typically starts around 24 $\mu$m, and
ends around 45 $\mu$m, outside of the \emph{Spitzer}/IRS wavelength
coverage, and it is carried by a solid state component. Its
identification remains unclear. \citet{Hony_02_MgS} argued that the
feature is due to MgS (magnesium sulfide), while rejecting other
carriers, such as HACs. The MgS
identification has been widely accepted by the community for the 30
$\mu$m feature seen in carbon-rich AGB stars, post-AGB stars and PNe 
\citep[e.g.][]{Zijlstra_06_LMC,Zhukovska_08_MgS,Woods_11_classification,Lombaert_12_LLPeg}. 
However,
\citet{Zhang_09_30mic} have demonstrated using energetic arguments
that the abundance of MgS is not sufficient to explain the strength of
the feature in post-AGB star HD56126, thus creating considerable
doubt concerning the validity of the MgS identification. In a recent
analysis of the 30 $\mu$m feature observed in a sample of extreme
carbon stars, \citet{Messenger_13_30mic} search for a correlation
between the 30 $\mu$m feature, and the 11.3 $\mu$m SiC (silicon
carbide) feature, on the grounds that the 30 $\mu$m feature is likely
carried by a carbide or sulfide component. They find that a
correlation exists within specifically this class of objects,
suggesting that the abundance of the carrier of the 30 $\mu$m feature
is linked to the SiC abundance. Analysis of laboratory spectra shows
that various types of sulfides, so not just MgS, all show a resonance
at approximately the right wavelength \citep{Messenger_13_30mic}.

In this work, however, we explore the possibility that the 30 $\mu$m
feature is due to the carrier of the dust continuum. The dust
continuum in carbon-rich AGB stars and PNe is generally thought to be
due to amorphous carbon \citep[e.g.][]{Srinivasan_10_Cstars}, but
there are hints that graphite is more appropriate
\citep[e.g.][]{Speck_09_graphite}. Comparing the opacities of
amorphous carbon and graphite shows that while amorphous carbon
produces a featureless continuum, graphite exhibits a very broad,
strong feature peaking around 40 $\mu$m, which shifts to longer
wavelengths for larger grains \citep{Draine_84_optprop}.  The
  position of the feature also seems to be a function of the
  conductivity of graphite, and \citet{Jiang_13_30um} show that
  for graphite with a d.c.~conductivity of 100 $\Omega^{-1}$ cm$^{-1}$
  the resonance is found at 30 $\mu$m.  Therefore, if we accept that
carbon in its crystalline form could produce a feature around 30-40
$\mu$m, and that the exact position of this feature may be dependent
the grain properties, such as size and shape, we can investigate
whether it is possible to construct a dust opacity that explains both
the long-wavelength dust continuum and the strength of the 30 $\mu$m feature
at the same time.

\subsubsection{Deriving the opacity function using IC418}
\label{sec:30micQ}

\begin{table}
\caption{Wavelength ranges blocked in the spectrum of IC418, prior to
continuum fitting. \label{tab:blockfeats}}
\begin{tabularx}{\columnwidth}{@{}XcYl@{}}
\hline\hline
{Blocked range ($\mu$m)} & {Comment}\\
\hline

       $<$13.5 & broad 11 $\mu$m and 11 $\mu$m PAH bands\\
       15.5-16 &[Ne\,{\sc iii}] 15.5 $\mu$m \\
       17-18 & 17.4 $\mu$m C$_{60}$ band and H~{\sc i} 17.61 $\mu$m\\
       18.5-19.5 &18.9 $\mu$m C$_{60}$ band and [S\,{\sc iii}] 18.7 $\mu$m \\
       23-45 & 30 $\mu$m feature \\
       51-53 & [O\,{\sc iii}] 51.8 $\mu$m\ \\
       56-58 & [N\,{\sc iii}] 57.3 $\mu$m\\
       62-64 & [O\,{\sc i}] 63.2 $\mu$m\\
       87-90 &[O\,{\sc iii}] 88.4 $\mu$m\\
       121-123 &  [N\,{\sc ii}] 121.8 $\mu$m\\
       144-147 &[O\,{\sc i}] 145.5 $\mu$m\\
       155-160 &[C\,{\sc ii}] 157 $\mu$m \\
\hline
\end{tabularx}
\end{table}

IC418 is the most suitable source to constrain the opacity function of
the carrier responsible for both the continuum and the 30 $\mu$m
feature, because it is the only source in our sample where
spectroscopy is available at $\lambda$ $>$40 $\mu$m, allowing us to
trace the full extent of the 30 $\mu$m feature, and have an additional
stretch of continuum baseline at even longer wavelengths. The
constructed opacity function can then be applied to the other objects
in our sample.

First, we removed all emission lines from the spectrum of IC418, by
blocking the data at the wavelength ranges listed in
Table~\ref{tab:blockfeats}. All data with $\lambda$$<$13.5 $\mu$m has
been blocked because of a change in slope due to contributions from
hot dust and PAH emission. The remaining spectrum was then re-binned
with a spectral resolution of $\Delta\lambda$/$\lambda$=10 to reduce
the noise and make the spectrum more smooth. All data points in the
re-binned spectrum of IC418 were assigned a flux uncertainty of 5\%,
in line with typical values for the \emph{Spitzer}/IRS and \emph{ISO}
flux calibration accuracy.

Assuming that dust in PNe is found to be present at a range of
temperatures, reflecting a range of distances to the central star, we
follow the method put forward by \citet{Kemper_02_NGC6302}. This model
assumes that dust is distributed around the central star following a
power-law $\rho(r) \propto r^{-p}$, with the temperature distribution
also follows a power-law $T(r) \propto r^{-q}$. Thus, following
\citet{Kemper_02_NGC6302} we can derive, for spherical grains, that
\begin{equation}
  F_\nu = C \cdot Q_{\mathrm{abs}}(\lambda) \int^{T_\mathrm{max}}_{T_\mathrm{min}} T^{-\alpha} B_\lambda (T) dT
\label{eq:dustem}
\end{equation}   
with $\alpha = (3-p)/q > 0$.
\noindent 
$Q_{\mathrm{abs}}(\lambda)$ is the absorption efficiency as a function of
wavelength. Relevant species that could contribute to the continuum
are amorphous carbon and graphite, as discussed before, but also
SiC. Both amorphous carbon and SiC essentially follow a power-law
$\propto \lambda^{-\beta}$ for $\lambda$$>$13.5 $\mu$m. Graphite does
deviate from this power-law behavior due to the strong resonance
around 40 $\mu$m which may be related to the 30 $\mu$m feature we are
trying to fit. This means we can substitute $\lambda^{-\beta}$ for
$Q_{\mathrm{abs}}(\lambda)$ in Equation ~(\ref{eq:dustem}), which yields
\begin{equation}
  F_\nu = C \cdot \lambda^{-\beta} \int^{T_\mathrm{max}}_{T_\mathrm{min}} T^{-\alpha} B_\lambda (T) dT
\label{eq:continuum}
\end{equation}   
\noindent This equation is fitted to the
re-binned spectral data of the continuum emission due to the dust in 
IC418 using the MPFIT algorithm \citep{Markwardt_09_MPFIT} excluding the
wavelengths listed in Table \ref{tab:blockfeats}. The fit is
not very sensitive to the minimum temperature at the outskirts of the
nebula, so we fix $T_{\mathrm{min}}$=20 K, a temperature more or
less characteristic for dust in the interstellar medium. This results
in only four free parameters, namely, the maximum dust temperature
$T_{\mathrm{max}}$ closest to the central star, the power-law index
$\alpha$, describing the physical properties of the nebula, the
power-law index $\beta$ describing the opacity function of the continuum
and the scaling factor $C$.

\subsubsection{Derivation of opacities}

Due to the presence of two power-laws in Equation ~(\ref{eq:continuum}) (one
over $T$, and one over $\lambda$) there is a certain amount of degeneracy
in the resulting fit.  Trial and error shows that $\alpha$ is
approximately equal to unity, so we have fixed it at that value, to
find that the index $\beta$ describing the opacity power-law, is 1.20. The
scaling factor $C$=3.3(--10), and we find that the maximum
temperature $T_{\mathrm{max}}$=155.3 K.

\begin{figure}
\includegraphics[clip,width=\columnwidth]{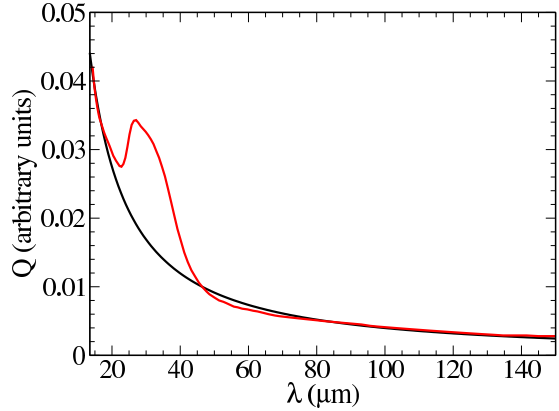}
\caption{The derived opacity for the carrier of the continuum and the
  30 $\mu$m feature (red), compared to a power-law opacity $\propto
  \lambda^{-1.2}$ (black). The $Q$-values are in arbitrary
  units. \label{fig:q30}}
\end{figure}

We can use these fit parameters now as true values in
Equation ~(\ref{eq:dustem}), while we leave $Q_{\mathrm{abs}}(\lambda)$ as an
unknown. Using the spectrum of IC418 with all emission lines and
features, except the 30 $\mu$m feature, removed as $F_\nu (\lambda)$,
we can derive $Q_{\mathrm{abs}}(\lambda)$.  The result is a table of Q
values over a range from 13.5 to $\sim$150 $\mu$m, which we have
re-binned with a spectral resolution of $\lambda/\Delta\lambda$=10.
The derived opacity for the 30 $\mu$m feature and the continuum
together is shown in Fig. \ref{fig:q30} with the red line, along with
a $\lambda^{-1.2}$ power-law in black. Outside the 30 $\mu$m feature,
the derived opacity follows the power-law quite closely, and is
therefore able to reproduce the continuum emission. Note that the
derived opacity is given in arbitrary units, because we did not
specify any grain size or shape.  The mass contained in this component
can therefore not be derived.

\subsubsection{Fitting the 30 $\mu$m feature in the entire sample}

 \begin{table}
 \caption{Fit parameters for all sources, for the solution to 
Equation~\ref{eq:dustem}, using the joint opacity presented in 
Fig. \ref{fig:q30} for $Q_{\mathrm{abs}}$.  \label{tab:30mufit}}
\begin{tabularx}{\columnwidth}{@{}XlYcYc@{}}
\hline\hline
{Nebula} & 
{$T_{\mathrm{max}}$(K)} & 
{$C$}\\
\hline
Hen2-68  & 139.1 $\pm$ 0.1& 2.57(--11)\\ 
IC418   & 146.1 $\pm$ 0.1 & 6.70(--10)\\
K3-62   & 138.5 $\pm$ 0.2 & 5.86(--11)\\
M1-6    & 150.3 $\pm$ 0.2 & 3.58(--11)\\
M1-9    & 147.0 $\pm$ 0.9  &5.91(--12)\\
M1-11   & 149.5 $\pm$ 0.1 & 2.00(--10)\\
M1-12   & 140.8 $\pm$ 0.1 & 4.77(--11)\\
M1-20   & 148.1 $\pm$ 0.1 & 1.33(--11)\\
SaSt2-3 & 148.9 $\pm$ 0.1 & 9.00(--13)\\
Tc1     & 117.2 $\pm$ 0.1 & 1.75(--10)\\
\hline
\end{tabularx}
\end{table}

\begin{figure*}
\centering
\resizebox{\hsize}{!}{
\includegraphics[clip]{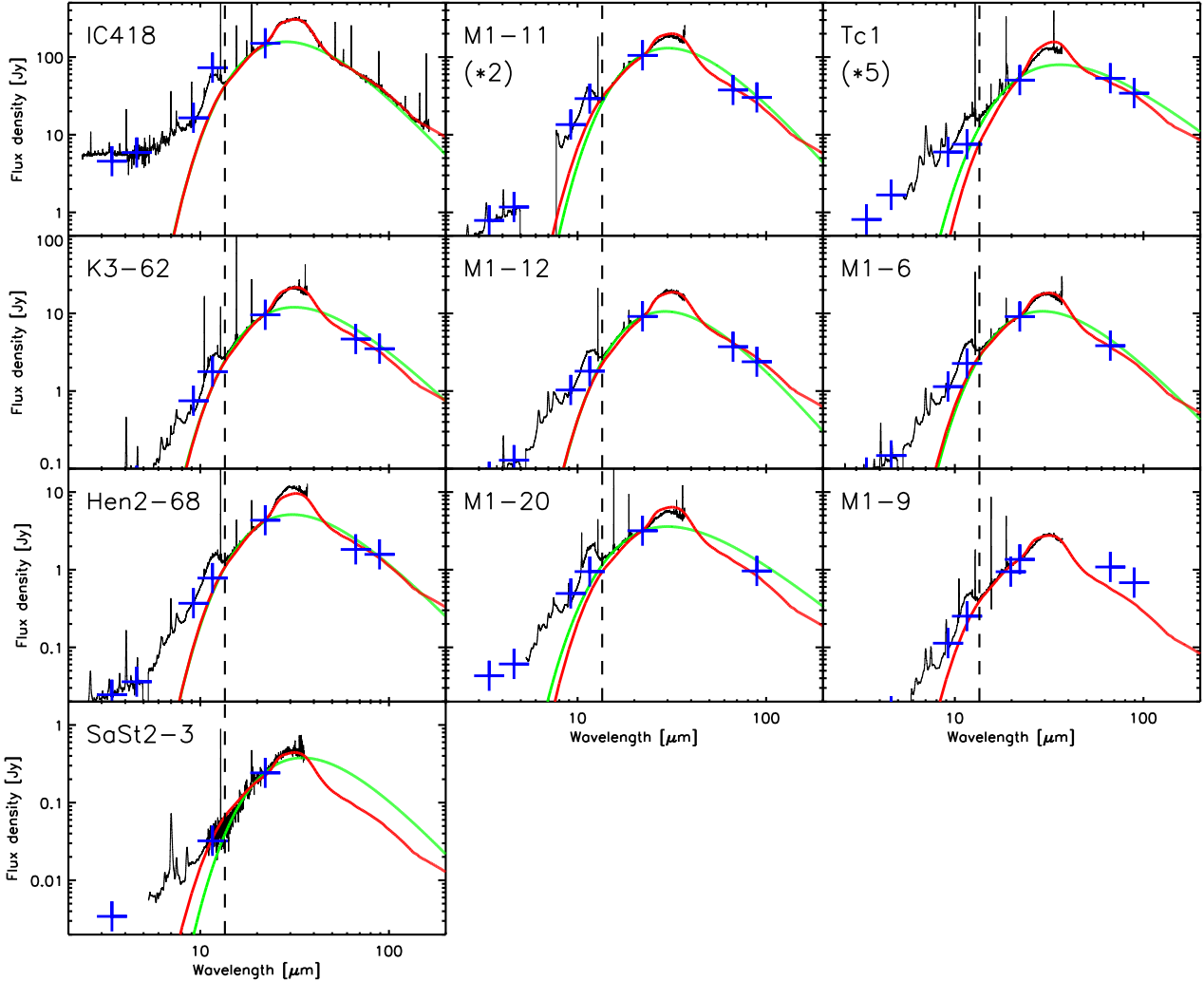}
}
\caption{
The best fit results for each of the objects to the continuum 
emission beyond 13.5 $\mu$m, and the 30 $\mu$m feature. 
The spectra are shown in black, and the photometry points with 
blue crosses. The red line shows the best fit result to Equation
 (\ref{eq:dustem}), which was fit to the data to the right of 
the dashed line, e.g. with $\lambda$$>$13.5 $\mu$m. The best 
fit modified blackbody, e.g. a single temperature blackbody 
multiplied by a $\lambda^{-p}$ emissivity law, to the same data 
is shown for comparison in each of the panels with a green line. 
For clarity, the spectra of M1-11 and Tc1 are shifted by 
the indicated factor. 
\label{fig:30umfit}}
\end{figure*}

The joint opacity $Q_{\mathrm{abs}} (\lambda)$, shown in
Fig. \ref{fig:q30}, can be used in Equation~(\ref{eq:dustem}) to fit the
continuum and 30 $\mu$m emission in all sources in the sample. Again,
we treat $T_{\mathrm{max}}$ and $C$ as free parameters. Experience
shows that $\alpha$ always has a value close to unity, so we fixed
this parameter at that value. The equation was solved using the MPFIT
routine \citep{Markwardt_09_MPFIT}, and the results for all targets in
our sample is shown in Table~\ref{tab:30mufit} and
Fig. \ref{fig:30umfit}. The first column of Table \ref{tab:30mufit} gives 
the source name, the second column the fitted $T_{\mathrm{max}}$, 
and the third column the scaling factor $C$. 
In Fig. \ref{fig:30umfit}, the spectra are 
shown in black, and the photometry points with blue crosses. 
The photometry comes from the four \emph{WISE} bands ($\lambda_{c}$=3.4, 
4.6, 11.6, and 22.1 $\mu$m) and the \emph{AKARI}/IRC \& FIS bands (9.2, 19.8, 66.7, and 89.2 $\mu$m). 
The red line shows the best fit result to Equation~(\ref{eq:dustem}), which was fit
 to the data to the right of the dashed line, e.g. with 
$\lambda$$>$13.5 $\mu$m. The best fit modified black-body, e.g. a single 
temperature black-body multiplied by a $\lambda^{-p}$ emissivity law, 
to the same data is shown for comparison 
with a green line. For clarity, the spectra of M1-11 and Tc1 are shifted by the indicated factor. 
The uncertainties on the derived maximum
temperature are of the order of 0.1--0.9 K, while the scaling factor is 
simply a linear factor applied to minimize the $\chi^2$ values, and hence
has an uncertainty of less than 1 $\%$.

Fig. \ref{fig:30umfit} shows that for all sources in our sample we can
achieve a good fit to the 30 $\mu$m feature and continuum at the same
time, using the opacity table derived from the full spectrum of IC418
(Fig. \ref{fig:q30}), and by applying the emission model described in
Equation~(\ref{eq:dustem}). Therefore, the fit results for our
  sample are consistent with having a single carrier for the continuum
  emission ($> 13.5$ $\mu$m) and the 30 $\mu$m feature.  The exact
composition of this material remains unknown at this time, but we
speculate that it may be related to HAC or graphite, because these
species are known to have features in this wavelength range
\citep{Grishko_01_HAC,Draine_84_optprop,Jiang_13_30um}.  The opacity
plot (Fig. \ref{fig:q30}) provides a starting point for future
attempts to identify the material responsible for the continuum and
the 30 $\mu$m feature.

\section{Discussion}
\subsection{Fullerenes in Galactic PNe}

\begin{figure}
\centering
\resizebox{\hsize}{!}{
\includegraphics[clip,bb=41 185 520 647]{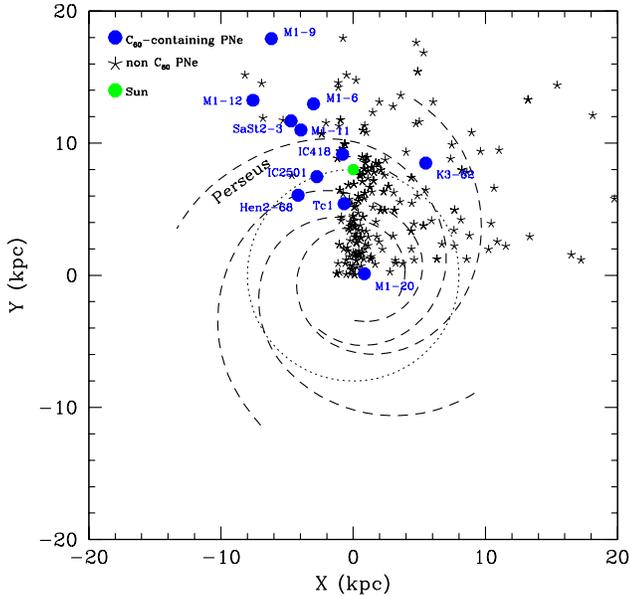}
}
\caption{The positions of the 254 non-C$_{60}$ (asterisks) and the 11
 C$_{60}$ PNe (filled blue circles) on a face-on map of the Milky Way. 
The position of the Sun is indicated by the filled green circle. The
 solar circle is indicated by the dotted line.
\label{fig:distpn}}
\end{figure}

\begin{figure} 
\centering
\resizebox{\hsize}{!}{
\includegraphics[clip,bb=50 274 512 590]{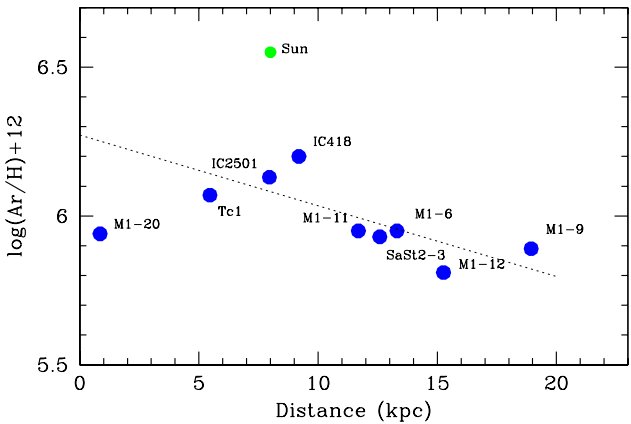}
}
\caption{The diagram between the distance from the Galactic center and
 the Ar abundances of our sample. The dotted line is the relation
 established among our sample,
 except for M1-20.   
\label{fig:argrad}}
\end{figure}

Out of 338 Galactic PNe observed with the 
\emph{Spitzer}/IRS spectroscopic modules, we found a total of eleven
objects in which C$_{60}$ is present using the SH data, 
six of which (Hen2-68, IC2501, K3-62, M1-6, M1-9, and SaSt2-3) are new
detections. These eleven sources
were identified by the presence of the 17.4 and 18.9 $\mu$m C$_{60}$
resonances. SaSt2-3 and Tc1 also
clearly show strong features at 7.0 and 8.5 $\mu$m due to
C$_{60}$. Except for SaSt2-3 and Tc1, the 8.5 $\mu$m C$_{60}$ band is
heavily contaminated by the 8.6 $\mu$m PAH band.

The large number of Galactic PNe checked for the presence of C$_{60}$
represent a considerable fraction of the total number of known (C-rich
and O-rich) PNe in the Milky Way \citep[$\sim$1200;][]{Acker:1992aa},
thus, from this representative sample, we may conclude that the
detectable presence of C$_{60}$ is rather rare, 
with a detection rate of one out of $\sim$30.7 PNe (=338/11).

Among our 338 PNe, the distances to 264 objects were measured by
\citet{Stanghellini:2010aa}. Using their results, 
we plot the positions of these PNe on a face-on map of the Milky Way
shown in Fig. \ref{fig:distpn}. The positions of the spiral arms are taken from
\citet{Cordes:2002aa}. \citet{Stanghellini:2010aa} did not measure the
distance to SaSt2-3, therefore we assumed the distance of 6 kpc, 
as earlier mentioned. Most PNe are concentrated to the Galactic
center, however, there are very few C$_{60}$-containing PNe 
located in that direction, only M1-20 (in the Galactic bulge) and Tc1. 
Most of the C$_{60}$-containing PNe are in the Galactic anti-center 
and they are predominantly located in the
region beyond the Perseus spiral arm.

In Fig. \ref{fig:argrad}, we plotted the diagram of the Galactocentric
distance against the Ar abundances of C$_{60}$-containing PNe. 
To investigate whether progenitor stars of C$_{60}$-containing PNe are 
initially metal-rich, we used Ar as a metallicity because Ar is not 
synthesized in low-mass AGB stars. The dotted line in Fig. \ref{fig:argrad} 
is the relation established between our sample, except for M1-20 and it is represented by 
\begin{equation}
{\rm Ar} = (-0.02\pm0.01)~{D_{gal}} + (6.27\pm0.10),
\end{equation}
\noindent where $D_{gal}$ is the Galactocentric distance in
kpc. The correlation factor is --0.48. Our Ar gradient (--0.02)
estimated by fitting to our sample is consistent with the result by 
\citet{2004AJ....127.2284H}, who investigated the Ar abundance gradient
using Galactic disk PN abundances. They reported that 
their correlation was 
(--0.03$\pm$0.01)${D_{gal}}$+(6.58$\pm$0.079). Their measured 
Ar abundances in PNe located around $D_{gal}$=5 kpc and 10 kpc seem 
to be from $\sim$6.3 to $\sim$6.8 dex and from $\sim$6 to $\sim$6.5 dex,
respectively. Therefore, we can conclude that C$_{60}$ PNe are
relatively metal-poor.

The O abundance amongst our sample can be represented by
\begin{equation}
{\rm O} = (-0.02\pm0.01)~{D_{gal}} + (8.64\pm0.10).
\end{equation}
\noindent The correlation factor is --0.54. \citet{2004AJ....127.2284H}
reported a slope of -0.037$\pm$0.008 and an intercept of 8.97$\pm$0.069.
We need to be careful when interpreting the value derived for the O abundance; because O is slightly 
increased by the helium-burning and the third dredge-up 
during the AGB phase, its abundance does not indicate the initial
metallicity. We can conclude that C$_{60}$ PNe are 
relatively metal-poor objects even if O is increased during the AGB phase.

If our predicted C abundances are correct, the C/O ratio
in our sample is increasing as $D_{gal}$ is increasing. Therefore, 
that is a reason why most the C$_{60}$ PNe are located outside the solar 
circle. To verify this, we need to measure the C/O ratio using the 
same type of emission-lines in the future.

\subsection{The nature of the carrier of the 11 $\mu$m band}
\begin{figure}
\centering
\resizebox{\hsize}{!}{
\includegraphics[bb=18 155 421 415,clip]{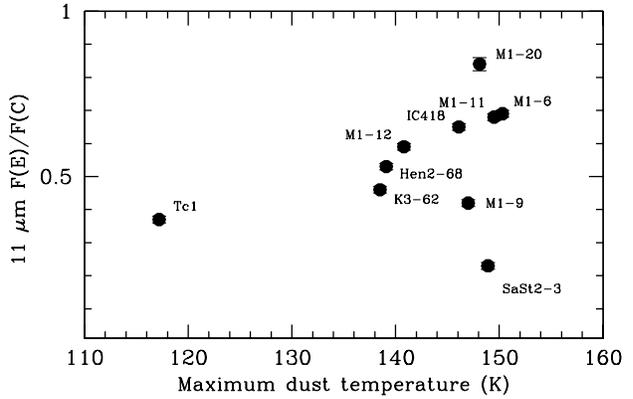}
}
\caption{\label{Fig:td11} 
The plot between the band strength of the broad 11 $\mu$m band
 and the maximum dust temperature $T_{\rm max}$.}
\end{figure}

We have shown in Section 4.7 that the 11 $\mu$m band emission does not
  correlate with the C$_{60}$ emission or the central star's effective
  temperature. However, when we plot $F$(E)/$F$(C) against the $T_{\rm
  max}$(dust) derived in Section 4.8, we find that 
  the 11 $\mu$m band strength feature seems to scale with the thermal
  dust emission, as shown in Fig. \ref{Fig:td11}, if we exclude the noisy
  spectrum of SaSt2-3. The correlation coefficients are 0.69 without
  SaSt2-3 and 0.39 with SaSt2-3, respectively. This correlation suggests
  that the carrier of this feature is a thermally heated dust component. 
  On the other hand, substructure in the 11 $\mu$m feature at 11.3 
  $\mu$m points to a carrier that is related to PAHs. 
  A similar broad feature is seen in Magellanic Cloud PNe,
  where it is thought to contain contributions from thermal emission
  from SiC and stochastically heated PAHs and PAH-like species
  \citep{Bernard-Salas_09_LMC}.

\subsection{Relative fraction of the 17.4 and 18.9 
$\mu$m C$_{60}$ band flux to the dust continuum flux}

\begin{table}
\caption{Integrated fluxes of the dust continuum in 10.2-200 $\mu$m and 
C$_{60}$ 17.4 and 18.9 $\mu$m emissions. \label{intdust}}
\begin{tabularx}{\columnwidth}{@{}XlYcYcYc@{}}
\hline\hline
{Nebula} & $F$(dust)                & $F$(C$_{60}$) & $F$(C$_{60}$)/$F$(dust)\\
         &(erg s$^{-1}$ cm$^{-2}$)   &(erg s$^{-1}$ cm$^{-2}$)&($\times$10$^{-2}$)\\
\hline
Hen2-68 & 8.87(--10) & 1.18(--12) & 0.13 \\ 
IC418 & 2.92(--8) & 6.48(--11) & 0.22 \\ 
K3-62 & 1.98(--9) & 1.92(--12) & 0.10 \\ 
M1-6 & 1.82(--9) & 2.93(--12) & 0.16 \\ 
M1-9 & 2.62(--10) & 6.34(--13) & 0.24 \\ 
M1-11 & 9.71(--9) & 1.05(--11) & 0.11 \\ 
M1-12 & 1.80(--9) & 3.50(--12) & 0.19 \\ 
M1-20 & 6.52(--10) & 1.31(--12) & 0.20 \\ 
SaSt2-3 & 4.06(--11) & 4.71(--13) & 1.16 \\ 
Tc1 & 2.71(--9) & 3.24(--11) & 1.20 \\ 
\hline
\end{tabularx}
\end{table}

By combining the best fits for the 13.5-200 $\mu$m SED presented
in Fig. \ref{fig:30umfit} and the spline function fits for the 5.3-23
$\mu$m continuum, we estimated the integrated dust continuum flux in 
10.2-200 $\mu$m. We estimated the integrated fluxes in 22.7-200 $\mu$m and in
10.2-22.7 $\mu$m by using the former and latter ones, respectively, and 
then we simply summed these fluxes. The integrated dust continuum
$F$(dust) is listed in the second column of Table \ref{intdust}. The
subsequent two columns are the total flux of the 17.4 and 18.9 $\mu$m
emissions $F$(C$_{60}$) and its fraction with respect to 
$F$(dust). The high $F$(C$_{60}$)/$F$(dust) also indicates that 
SaSt2-3 and Tc1 are unusually C$_{60}$-rich.

\section{Summary}
We examined the properties of the nebulae, the central stars, and 
the infrared spectroscopic features in 
eleven Galactic C$_{60}$-bearing PNe based on ground-based and space
telescope data, and theoretical models. Six of the sources in our sample (Hen2-68, IC2501,
K3-62, M1-6, M1-9, and SaSt2-3) are newly identified as C$_{60}$-containing PNe.
Photo-ionization models indicate that the nebular 
chemical abundance pattern and the current evolutionary status of the 
central stars are very similar, strongly suggesting that these
C$_{60}$-bearing PNe have evolved from progenitors with similar initial mass 
and chemical composition. We found that the chemical abundances of
C$_{60}$-containing PNe can be explained by AGB nucleosynthesis 
models for initially 1.5-2.5 M$_{\odot}$ stars with $Z$=0.004. Their
 metallicity suggests that the progenitors are from an older
 population. We plotted the locations of the C$_{60}$-containing PNe on a
 face-on map of the Milky Way and we found that most of these PNe are
 outside the solar circle, consistent with low metallicity values. 
The effective temperatures
differ from source to source, ranging from 29\,750 to 51\,650 K. 
There is no common characteristic in the nebular shape. Among the 
eleven PNe, SaSt2-3 and Tc1 
clearly show strong C$_{60}$ bands at 7.0 and 8.5 $\mu$m, while all
eleven objects exhibit the 17.4 and 18.9 $\mu$m C$_{60}$ features. The flux
ratio between the 17.4 and 18.9 $\mu$m C$_{60}$ feature is rather
constant amongst the sample, with an average value of 0.49. 
The PAH profile over 6-9 $\mu$m in our sample 
is of the more chemically-processed class A. 
We estimated the number of C-atoms in typical PAHs using
the ratio of the 3.3 $\mu$m to the 11.3 $\mu$m PAH flux. The number of
C-atoms per PAH in C$_{60}$-containing PNe is small, compared to
that in non-C$_{60}$ PNe.
We studied the nature of the 11 and 30 $\mu$m features. Both are showing to be due to thermal dust emission. 
The 11 $\mu$m band may have a contribution due to stochastically heated
PAHs, while the carrier of the 30 $\mu$m feature is shown to be one
and the same as the carrier of the dust continuum.

\section*{Acknowledgments}
We are grateful to the anonymous referee for the useful suggestions which
 greatly improved this article. 
We thank Elisabetta Micelotta and Anthony Jones for numerous fruitful
discussions. MO thanks the astronomy group at the University of Western Ontario for
their kind support during his visit. FK acknowledges support from the
National Science Council in the form of grant
NSC100-2112-M-001-023-MY3. JBS wishes to
  acknowledge the support from a Marie Curie Intra-European Fellowship
  within the 7th European Community Framework Program under project
  number 272820. JC and EP acknowledge support from an NSERC
Discovery Grant and a startup grant from the Department of Physics and
Astronomy at Western University. This work is based on archival data
obtained with the Spitzer Space Telescope, which is operated by the
Jet Propulsion Laboratory, California Institute of Technology under a
contract with NASA. Support for this work was provided by an award
issued by JPL/Caltech.  This work is in part based on \emph{HST}
archive data from the CADC and ESO archive data.  This work is in part
based on observations with \emph{AKARI}, a JAXA project with the
participation of ESA. This research has made grateful use of the
SIMBAD database, operated at CDS, Strasbourg, France; and of NASA's
Astrophysics Data System Bibliographic Services.

\appendix
\section{Flux measurements of the 7.0 and 8.5 microns C$_{60}$ band}
As mentioned in Sections \ref{akari_pah} and \ref{5-9pah}, all C$_{60}$-containing
PNe in our sample show PAH features. Therefore, we assumed that 
the emission line around 8.5 $\mu$m is not only due to C$_{60}$, but
also shows a contribution due to the 8.6 $\mu$m PAH band. 
We measured the C$_{60}$ band 
in SaSt2-3 and Tc1 using a single Gaussian, thus obtaining the FWHM of
the 8.5 $\mu$m C$_{60}$ band. In a multiple Gaussian fit to the 8.5 $\mu$m
C$_{60}$ and 8.6 $\mu$m PAH complex in PNe except for SaSt2-3 and Tc1, we adopted the
average FWHM of the 8.5 $\mu$m C$_{60}$ in SaSt2-3 and Tc1 ($\sim$0.18
$\mu$m) as the first assumption for the 8.5 $\mu$m C$_{60}$
component. We fix the FWHM of the 8.6 $\mu$m PAH band at the value of 0.23 $\mu$m measured 
from the \emph{ISO}/SWS spectrum of NGC7027 displayed in
Fig. \ref{Fig:spec-5-9um}b.

Measurements of the 7.0 $\mu$m C$_{60}$ band are more difficult. To 
investigate the excitation mechanism of C$_{60}$, flux measurements of
the 7.0 $\mu$m C$_{60}$ bands with any contamination subtracted are
necessary, as discussed by \citet{Jero:C60excitation}. The 7.0 $\mu$m
C$_{60}$ band 
is mainly contaminated by the 
[Ar\,{\sc ii}] 6.99 $\mu$m line. We do not
detect any pure rotational H$_{2}$ lines in any of the spectra, so we assume that the contribution from the 
H$_{2}$ $v$=0-0 S(5) 6.91 $\mu$m line is
negligibly small. In NGC7027, the flux ratio of the 6.92 $\mu$m
aliphatic band to the 6.2 $\mu$m PAH band is 0.12.

We estimated the flux in the 7.0 $\mu$m C$_{60}$ resonance by
deconstructing the observed 7.0 $\mu$m band in its components due to the
6.92 $\mu$m aliphatic band, the [Ar\,{\sc ii}] 6.99 $\mu$m line, and the 7.0
$\mu$m C$_{60}$ combination.
We assumed that the flux ratio of the 6.92 $\mu$m to
6.2 $\mu$m PAH bands (0.12) is the same as in NGC7027 and the 
flux ratio of [Ar\,{\sc ii}] 6.99 $\mu$m to [Ar\,{\sc iii}] 8.99 $\mu$m
derived with {\sc CLOUDY} (See Table \ref{Table:properties}). The predicted
ratio of $F$([Ar\,{\sc ii}] 6.99 $\mu$m)/$F$([Ar\,{\sc iii}] 8.99
$\mu$m) is in the last column. The values in Hen2-68 and K3-62 are estimated using a relation between the ratio and $T_{\rm eff}$ among our PNe with $>$36\,000 K;
\begin{eqnarray}
F([{\rm Ar\,~II}]~6.99~{\mu}{\rm m})/F([{\rm Ar\,~III}]~8.99~{\mu}{\rm m}) = \nonumber\\
-1.35\times10^{-4}T_{\rm eff}+6.16.
\end{eqnarray}
\noindent 
The predicted line ratio, in combination with the measured flux for the 
[Ar\,{\sc iii}] 8.99 $\mu$m line, was used to estimate the
contamination from [Ar\,{\sc ii}] 6.99 $\mu$m line to C$_{60}$ 7.0 $\mu$m feature. 

The total flux of the 7.0 $\mu$m complex ($F_{tot.}$(7.0$\mu$m)) is
simply the sum of the [Ar\,{\sc ii}] 6.99 $\mu$m 
($F$(6.99 $\mu$m)), the 6.92 $\mu$m aliphatic band ($F$(6.92 $\mu$m)),
and the C$_{60}$ 
7.0 $\mu$m band ($F_{{\rm C_{60}}}$(7.0 $\mu$m)).
Thus, the approximate flux of the 7.0 $\mu$m C$_{60}$ line can be written as 
\begin{eqnarray}
\lefteqn{F_{{\rm C_{60}}}(7.0~{\mu}{\rm m}) = F_{\rm tot.}(7.0~{\mu}{\rm m})-} \nonumber\\
&& {\rm Const.}{\times}F(8.99~{\mu}{\rm m}) - 0.12~F(6.2~{\mu}{\rm m}),
\end{eqnarray}
where the constant is the flux ratio $F$([Ar\,{\sc ii}] 6.99
$\mu$m)/$F$([Ar\,{\sc iii}] 8.99 $\mu$m) and $F$(6.2 $\mu$m) 
is the 6.2 $\mu$m PAH flux.

\label{lastpage}

\begin{thebibliography}{74}
\expandafter\ifx\csname natexlab\endcsname\relax\def\natexlab#1{#1}\fi

\bibitem[{{Acker} {et~al.}(1992){Acker}, {Marcout}, {Ochsenbein}, {Stenholm},
  {Tylenda}, \& {Schohn}}]{Acker:1992aa}
{Acker} A., {Marcout} J., {Ochsenbein} F., {Stenholm} B., {Tylenda} R.,
  {Schohn} C., 1992, {The Strasbourg-ESO Catalogue of Galactic Planetary
  Nebulae. Parts I, II.} Observatoire de Strasbourg

\bibitem[{{Acker} \& {Neiner}(2003)}]{Acker:2003aa}
{Acker} A., {Neiner} C., 2003, \aap, 403, 659

\bibitem[{{Allamandola} {et~al.}(1989){Allamandola}, {Tielens}, \&
  {Barker}}]{Allamandola:1989aa}
{Allamandola} L.~J., {Tielens} A.~G.~G.~M., {Barker} J.~R., 1989, \apjs, 71,
  733

\bibitem[{{Bernard-Salas} {et~al.}(2012){Bernard-Salas}, {Cami}, {Peeters},
  {Jones}, {Micelotta}, \& {Groenewegen}}]{Jero:C60excitation}
{Bernard-Salas} J., {Cami} J., {Peeters} E., {Jones} A.~P., {Micelotta} E.~R.,
  {Groenewegen} M.~A.~T., 2012, \apj, 757, 41

\bibitem[{{Bernard-Salas} {et~al.}(2009){Bernard-Salas}, {Peeters}, {Sloan},
  {Gutenkunst}, {Matsuura}, {Tielens}, {Zijlstra}, \&
  {Houck}}]{Bernard-Salas_09_LMC}
{Bernard-Salas} J., {Peeters} E., {Sloan} G.~C., {Gutenkunst} S., {Matsuura}
  M., {Tielens} A.~G.~G.~M., {Zijlstra} A.~A., {Houck} J.~R., 2009, \apj, 699,
  1541

\bibitem[{{Boersma} {et~al.}(2008){Boersma}, {Bouwman}, {Lahuis}, {van
  Kerckhoven}, {Tielens}, {Waters}, \& {Henning}}]{Boersma:2008aa}
{Boersma} C., {Bouwman} J., {Lahuis} F., {van Kerckhoven} C., {Tielens}
  A.~G.~G.~M., {Waters} L.~B.~F.~M., {Henning} T., 2008, \aap, 484, 241

\bibitem[{{Cahn} {et~al.}(1992){Cahn}, {Kaler}, \&
  {Stanghellini}}]{1992A&AS...94..399C}
{Cahn} J.~H., {Kaler} J.~B., {Stanghellini} L., 1992, \aaps, 94, 399

\bibitem[{{Cami} {et~al.}(2010){Cami}, {Bernard-Salas}, {Peeters}, \&
  {Malek}}]{Cami_2010}
{Cami} J., {Bernard-Salas} J., {Peeters} E., {Malek} S.~E., 2010, Science, 329,
  1180

\bibitem[{{Clayton} {et~al.}(2011){Clayton}, {De Marco}, {Whitney}, {Babler},
  {Gallagher}, {Nordhaus}, {Speck}, {Wolff}, {Freeman}, {Camp}, {Lawson},
  {Roman-Duval}, {Misselt}, {Meade}, {Sonneborn}, {Matsuura}, \&
  {Meixner}}]{2011AJ....142...54C}
{Clayton} G.~C., {De Marco} O., {Whitney} B.~A., {Babler} B., {Gallagher}
  J.~S., {Nordhaus} J., {Speck} A.~K., {Wolff} M.~J., {Freeman} W.~R., {Camp}
  K.~A., {Lawson} W.~A., {Roman-Duval} J., {Misselt} K.~A., {Meade} M.,
  {Sonneborn} G., {Matsuura} M., {Meixner} M., 2011, \aj, 142, 54

\bibitem[{{Cordes} \& {Lazio}(2002)}]{Cordes:2002aa}
{Cordes} J.~M., {Lazio} T.~J.~W., 2002, ArXiv Astrophysics e-prints

\bibitem[{{Dopita} \& {Hua}(1997)}]{Dopita:1997aa}
{Dopita} M.~A., {Hua} C.~T., 1997, \apjs, 108, 515

\bibitem[{Draine \& Lee(1984)}]{Draine_84_optprop}
Draine B.~T., Lee H.~M., 1984, \apj, 285, 89

\bibitem[{{Draine} \& {Li}(2007)}]{Draine:2007aa}
{Draine} B.~T., {Li} A., 2007, \apj, 657, 810

\bibitem[{{Evans} {et~al.}(2012){Evans}, {van Loon}, {Woodward}, {Gehrz},
  {Clayton}, {Helton}, {Rushton}, {Eyres}, {Krautter}, {Starrfield}, \&
  {Wagner}}]{2012MNRAS.421L..92E}
{Evans} A., {van Loon} J.~T., {Woodward} C.~E., {Gehrz} R.~D., {Clayton} G.~C.,
  {Helton} L.~A., {Rushton} M.~T., {Eyres} S.~P.~S., {Krautter} J.,
  {Starrfield} S., {Wagner} R.~M., 2012, \mnras, 421, L92

\bibitem[{{Ferland} {et~al.}(1998){Ferland}, {Korista}, {Verner}, {Ferguson},
  {Kingdon}, \& {Verner}}]{1998PASP..110..761F}
{Ferland} G.~J., {Korista} K.~T., {Verner} D.~A., {Ferguson} J.~W., {Kingdon}
  J.~B., {Verner} E.~M., 1998, \pasp, 110, 761

\bibitem[{{Garc{\'{\i}}a-Hern{\'a}ndez}
  {et~al.}(2011{\natexlab{a}}){Garc{\'{\i}}a-Hern{\'a}ndez}, {Iglesias-Groth},
  {Acosta-Pulido}, {Manchado}, {Garc{\'{\i}}a-Lario}, {Stanghellini},
  {Villaver}, {Shaw}, \& {Cataldo}}]{2011ApJ...737L..30G}
{Garc{\'{\i}}a-Hern{\'a}ndez} D.~A., {Iglesias-Groth} S., {Acosta-Pulido}
  J.~A., {Manchado} A., {Garc{\'{\i}}a-Lario} P., {Stanghellini} L., {Villaver}
  E., {Shaw} R.~A., {Cataldo} F., 2011{\natexlab{a}}, \apjl, 737, L30

\bibitem[{{Garc{\'{\i}}a-Hern{\'a}ndez}
  {et~al.}(2011{\natexlab{b}}){Garc{\'{\i}}a-Hern{\'a}ndez}, {Kameswara Rao},
  \& {Lambert}}]{Gar_2011}
{Garc{\'{\i}}a-Hern{\'a}ndez} D.~A., {Kameswara Rao} N., {Lambert} D.~L.,
  2011{\natexlab{b}}, \apj, 729, 126

\bibitem[{{Garc{\'{\i}}a-Hern{\'a}ndez}
  {et~al.}(2010){Garc{\'{\i}}a-Hern{\'a}ndez}, {Manchado},
  {Garc{\'{\i}}a-Lario}, {Stanghellini}, {Villaver}, {Shaw}, {Szczerba}, \&
  {Perea-Calder{\'o}n}}]{2010ApJ...724L..39G}
{Garc{\'{\i}}a-Hern{\'a}ndez} D.~A., {Manchado} A., {Garc{\'{\i}}a-Lario} P.,
  {Stanghellini} L., {Villaver} E., {Shaw} R.~A., {Szczerba} R.,
  {Perea-Calder{\'o}n} J.~V., 2010, \apjl, 724, L39

\bibitem[{{Gielen} {et~al.}(2011){Gielen}, {Cami}, {Bouwman}, {Peeters}, \&
  {Min}}]{2011A&A...536A..54G}
{Gielen} C., {Cami} J., {Bouwman} J., {Peeters} E., {Min} M., 2011, \aap, 536,
  A54

\bibitem[{{Grishko} {et~al.}(2001){Grishko}, {Tereszchuk}, {Duley}, \&
  {Bernath}}]{Grishko_01_HAC}
{Grishko} V.~I., {Tereszchuk} K., {Duley} W.~W., {Bernath} P., 2001, \apjl,
  558, L129

\bibitem[{{Henry} {et~al.}(2004){Henry}, {Kwitter}, \&
  {Balick}}]{2004AJ....127.2284H}
{Henry} R.~B.~C., {Kwitter} K.~B., {Balick} B., 2004, \aj, 127, 2284

\bibitem[{{Henry} {et~al.}(2010){Henry}, {Kwitter}, {Jaskot}, {Balick},
  {Morrison}, \& {Milingo}}]{Henry:2010aa}
{Henry} R.~B.~C., {Kwitter} K.~B., {Jaskot} A.~E., {Balick} B., {Morrison}
  M.~A., {Milingo} J.~B., 2010, \apj, 724, 748

\bibitem[{{Higdon} {et~al.}(2004){Higdon}, {Devost}, {Higdon}, {Brandl},
  {Houck}, {Hall}, {Barry}, {Charmandaris}, {Smith}, {Sloan}, \&
  {Green}}]{Higdon_2004}
{Higdon} S.~J.~U., {Devost} D., {Higdon} J.~L., {Brandl} B.~R., {Houck} J.~R.,
  {Hall} P., {Barry} D., {Charmandaris} V., {Smith} J.~D.~T., {Sloan} G.~C.,
  {Green} J., 2004, \pasp, 116, 975

\bibitem[{Hony {et~al.}(2002)Hony, Waters, \& Tielens}]{Hony_02_MgS}
Hony S., Waters L. B. F.~M., Tielens A. G. G.~M., 2002, \aap, 390, 533

\bibitem[{{Houck} {et~al.}(2004){Houck}, {Roellig}, {van Cleve}, {Forrest},
  {Herter}, {Lawrence}, {Matthews}, {Reitsema}, {Soifer}, {Watson}, {Weedman},
  {Huisjen}, {Troeltzsch}, {Barry}, {Bernard-Salas}, {Blacken}, {Brandl},
  {Charmandaris}, {Devost}, {Gull}, {Hall}, {Henderson}, {Higdon}, {Pirger},
  {Schoenwald}, {Sloan}, {Uchida}, {Appleton}, {Armus}, {Burgdorf},
  {Fajardo-Acosta}, {Grillmair}, {Ingalls}, {Morris}, \&
  {Teplitz}}]{Houck:2004aa}
{Houck} J.~R., {Roellig} T.~L., {van Cleve} J., {Forrest} W.~J., {Herter} T.,
  {Lawrence} C.~R., {Matthews} K., {Reitsema} H.~J., {Soifer} B.~T., {Watson}
  D.~M., {Weedman} D., {Huisjen} M., {Troeltzsch} J., {Barry} D.~J.,
  {Bernard-Salas} J., {Blacken} C.~E., {Brandl} B.~R., {Charmandaris} V.,
  {Devost} D., {Gull} G.~E., {Hall} P., {Henderson} C.~P., {Higdon} S.~J.~U.,
  {Pirger} B.~E., {Schoenwald} J., {Sloan} G.~C., {Uchida} K.~I., {Appleton}
  P.~N., {Armus} L., {Burgdorf} M.~J., {Fajardo-Acosta} S.~B., {Grillmair}
  C.~J., {Ingalls} J.~G., {Morris} P.~W., {Teplitz} H.~I., 2004, \apjs, 154, 18

\bibitem[{{Hubeny} \& {Lanz}(1995)}]{Hubeny:1995aa}
{Hubeny} I., {Lanz} T., 1995, \apj, 439, 875

\bibitem[{Jiang {et~al.}(2013)Jiang, Li, Liu, \& J.~Gao}]{Jiang_13_30um}
Jiang B., Li A., Liu J., J.~Gao A.~M., 2013, in IAU Symposium, Vol. 297,
  Diffuse Interstellar Bands, Cami J., Cox N., eds.

\bibitem[{{Karakas}(2010)}]{Karakas:2010aa}
{Karakas} A.~I., 2010, \mnras, 403, 1413

\bibitem[{Kemper {et~al.}(2002)Kemper, Molster, J\"ager, \&
  Waters}]{Kemper_02_NGC6302}
Kemper F., Molster F.~J., J\"ager C., Waters L. B. F.~M., 2002, \aap, 394, 679

\bibitem[{{Kondratyeva}(2003)}]{Kondratyeva:2003aa}
{Kondratyeva} L.~N., 2003, Astronomical and Astrophysical Transactions, 22, 181

\bibitem[{{Kroto} {et~al.}(1985){Kroto}, {Heath}, {Obrien}, {Curl}, \&
  {Smalley}}]{Kroto:1985aa}
{Kroto} H.~W., {Heath} J.~R., {Obrien} S.~C., {Curl} R.~F., {Smalley} R.~E.,
  1985, \nat, 318, 162

\bibitem[{{Lodders}(2003)}]{Lodders:2003aa}
{Lodders} K., 2003, \apj, 591, 1220

\bibitem[{{Lombaert} {et~al.}(2012){Lombaert}, {de Vries}, {de Koter}, {Decin},
  {Min}, {Smolders}, {Mutschke}, \& {Waters}}]{Lombaert_12_LLPeg}
{Lombaert} R., {de Vries} B.~L., {de Koter} A., {Decin} L., {Min} M.,
  {Smolders} K., {Mutschke} H., {Waters} L.~B.~F.~M., 2012, \aap, 544, L18

\bibitem[{{Markwardt}(2009)}]{Markwardt_09_MPFIT}
{Markwardt} C.~B., 2009, in Astronomical Society of the Pacific Conference
  Series, Vol. 411, Astronomical Data Analysis Software and Systems XVIII,
  {Bohlender} D.~A., {Durand} D., {Dowler} P., eds., p. 251

\bibitem[{{Messenger} {et~al.}(2013){Messenger}, {Speck}, \&
  {Volk}}]{Messenger_13_30mic}
{Messenger} S.~J., {Speck} A., {Volk} K., 2013, \apj, 764, 142

\bibitem[{{Micelotta} {et~al.}(2012){Micelotta}, {Jones}, {Cami}, {Peeters},
  {Bernard-Salas}, \& {Fanchini}}]{Micelotta:2012aa}
{Micelotta} E.~R., {Jones} A.~P., {Cami} J., {Peeters} E., {Bernard-Salas} J.,
  {Fanchini} G., 2012, \apj, 761, 35

\bibitem[{{Morisset} \& {Georgiev}(2009)}]{Morisset:2009aa}
{Morisset} C., {Georgiev} L., 2009, \aap, 507, 1517

\bibitem[{{Morisset} {et~al.}(2012){Morisset}, {Szczerba}, {Anibal
  Garc{\'{\i}}a-Hern{\'a}ndez}, \& {Garc{\'{\i}}a-Lario}}]{Morisset_2012}
{Morisset} C., {Szczerba} R., {Anibal Garc{\'{\i}}a-Hern{\'a}ndez} D.,
  {Garc{\'{\i}}a-Lario} P., 2012, in IAU Symposium, Vol. 283, IAU Symposium,
  pp. 452--453

\bibitem[{{Murakami} {et~al.}(2007){Murakami}, {Baba}, {Barthel}, {Clements},
  {Cohen}, {Doi}, {Enya}, {Figueredo}, {Fujishiro}, {Fujiwara}, {Fujiwara},
  {Garcia-Lario}, {Goto}, {Hasegawa}, {Hibi}, {Hirao}, {Hiromoto}, {Hong},
  {Imai}, {Ishigaki}, {Ishiguro}, {Ishihara}, {Ita}, {Jeong}, {Jeong},
  {Kaneda}, {Kataza}, {Kawada}, {Kawai}, {Kawamura}, {Kessler}, {Kester},
  {Kii}, {Kim}, {Kim}, {Kobayashi}, {Koo}, {Kwon}, {Lee}, {Lorente}, {Makiuti},
  {Matsuhara}, {Matsumoto}, {Matsuo}, {Matsuura}, {M{\"u}ller}, {Murakami},
  {Nagata}, {Nakagawa}, {Naoi}, {Narita}, {Noda}, {Oh}, {Ohnishi}, {Ohyama},
  {Okada}, {Okuda}, {Oliver}, {Onaka}, {Ootsubo}, {Oyabu}, {Pak}, {Park},
  {Pearson}, {Rowan-Robinson}, {Saito}, {Sakon}, {Salama}, {Sato}, {Savage},
  {Serjeant}, {Shibai}, {Shirahata}, {Sohn}, {Suzuki}, {Takagi}, {Takahashi},
  {Tanab{\'e}}, {Takeuchi}, {Takita}, {Thomson}, {Uemizu}, {Ueno}, {Usui},
  {Verdugo}, {Wada}, {Wang}, {Watabe}, {Watarai}, {White}, {Yamamura},
  {Yamauchi}, \& {Yasuda}}]{Murakami_2007}
{Murakami} H., {Baba} H., {Barthel} P., {Clements} D.~L., {Cohen} M., {Doi} Y.,
  {Enya} K., {Figueredo} E., {Fujishiro} N., {Fujiwara} H., {Fujiwara} M.,
  {Garcia-Lario} P., {Goto} T., {Hasegawa} S., {Hibi} Y., {Hirao} T.,
  {Hiromoto} N., {Hong} S.~S., {Imai} K., {Ishigaki} M., {Ishiguro} M.,
  {Ishihara} D., {Ita} Y., {Jeong} W.-S., {Jeong} K.~S., {Kaneda} H., {Kataza}
  H., {Kawada} M., {Kawai} T., {Kawamura} A., {Kessler} M.~F., {Kester} D.,
  {Kii} T., {Kim} D.~C., {Kim} W., {Kobayashi} H., {Koo} B.~C., {Kwon} S.~M.,
  {Lee} H.~M., {Lorente} R., {Makiuti} S., {Matsuhara} H., {Matsumoto} T.,
  {Matsuo} H., {Matsuura} S., {M{\"u}ller} T.~G., {Murakami} N., {Nagata} H.,
  {Nakagawa} T., {Naoi} T., {Narita} M., {Noda} M., {Oh} S.~H., {Ohnishi} A.,
  {Ohyama} Y., {Okada} Y., {Okuda} H., {Oliver} S., {Onaka} T., {Ootsubo} T.,
  {Oyabu} S., {Pak} S., {Park} Y.-S., {Pearson} C.~P., {Rowan-Robinson} M.,
  {Saito} T., {Sakon} I., {Salama} A., {Sato} S., {Savage} R.~S., {Serjeant}
  S., {Shibai} H., {Shirahata} M., {Sohn} J., {Suzuki} T., {Takagi} T.,
  {Takahashi} H., {Tanab{\'e}} T., {Takeuchi} T.~T., {Takita} S., {Thomson} M.,
  {Uemizu} K., {Ueno} M., {Usui} F., {Verdugo} E., {Wada} T., {Wang} L.,
  {Watabe} T., {Watarai} H., {White} G.~J., {Yamamura} I., {Yamauchi} C.,
  {Yasuda} A., 2007, \pasj, 59, 369

\bibitem[{{Onaka} {et~al.}(2007){Onaka}, {Matsuhara}, {Wada}, {Fujishiro},
  {Fujiwara}, {Ishigaki}, {Ishihara}, {Ita}, {Kataza}, {Kim}, {Matsumoto},
  {Murakami}, {Ohyama}, {Oyabu}, {Sakon}, {Tanab{\'e}}, {Takagi}, {Uemizu},
  {Ueno}, {Usui}, {Watarai}, {Cohen}, {Enya}, {Ootsubo}, {Pearson}, {Takeyama},
  {Yamamuro}, \& {Ikeda}}]{Onaka_2007}
{Onaka} T., {Matsuhara} H., {Wada} T., {Fujishiro} N., {Fujiwara} H.,
  {Ishigaki} M., {Ishihara} D., {Ita} Y., {Kataza} H., {Kim} W., {Matsumoto}
  T., {Murakami} H., {Ohyama} Y., {Oyabu} S., {Sakon} I., {Tanab{\'e}} T.,
  {Takagi} T., {Uemizu} K., {Ueno} M., {Usui} F., {Watarai} H., {Cohen} M.,
  {Enya} K., {Ootsubo} T., {Pearson} C.~P., {Takeyama} N., {Yamamuro} T.,
  {Ikeda} Y., 2007, \pasj, 59, 401

\bibitem[{{Otsuka} {et~al.}(2013){Otsuka}, {Kemper}, {Hyung}, {Sargent},
  {Meixner}, {Tajitsu}, \& {Yanagisawa}}]{Otsuka_2013}
{Otsuka} M., {Kemper} F., {Hyung} S., {Sargent} B.~A., {Meixner} M., {Tajitsu}
  A., {Yanagisawa} K., 2013, \apj, 764, 77

\bibitem[{{Otsuka} {et~al.}(2010){Otsuka}, {Tajitsu}, {Hyung}, \&
  {Izumiura}}]{Otsuka:2010aa}
{Otsuka} M., {Tajitsu} A., {Hyung} S., {Izumiura} H., 2010, \apj, 723, 658

\bibitem[{{Patriarchi} {et~al.}(1989){Patriarchi}, {Cerruti-Sola}, \&
  {Perinotto}}]{1989ApJ...345..327P}
{Patriarchi} P., {Cerruti-Sola} M., {Perinotto} M., 1989, \apj, 345, 327

\bibitem[{{Peeters} {et~al.}(2002){Peeters}, {Hony}, {Van Kerckhoven},
  {Tielens}, {Allamandola}, {Hudgins}, \& {Bauschlicher}}]{Peeters02}
{Peeters} E., {Hony} S., {Van Kerckhoven} C., {Tielens} A.~G.~G.~M.,
  {Allamandola} L.~J., {Hudgins} D.~M., {Bauschlicher} C.~W., 2002, \aap, 390,
  1089

\bibitem[{{Peeters} {et~al.}(2012){Peeters}, {Tielens}, {Allamandola}, \&
  {Wolfire}}]{Peeters:2012aa}
{Peeters} E., {Tielens} A.~G.~G.~M., {Allamandola} L.~J., {Wolfire} M.~G.,
  2012, \apj, 747, 44

\bibitem[{{Pereira} \& {Miranda}(2007)}]{Pereira:2007aa}
{Pereira} C.-B., {Miranda} L.-F., 2007, \aap, 467, 1249

\bibitem[{{Pottasch} \& {Bernard-Salas}(2006)}]{Pottasch:2006aa}
{Pottasch} S.~R., {Bernard-Salas} J., 2006, \aap, 457, 189

\bibitem[{{Pottasch} {et~al.}(2011){Pottasch}, {Surendiranath}, \&
  {Bernard-Salas}}]{Pottasch:2011aa}
{Pottasch} S.~R., {Surendiranath} R., {Bernard-Salas} J., 2011, \aap, 531, A23

\bibitem[{{Preite-Martinez} {et~al.}(1989){Preite-Martinez}, {Acker},
  {Koeppen}, \& {Stenholm}}]{Preite-Martinez:1989aa}
{Preite-Martinez} A., {Acker} A., {Koeppen} J., {Stenholm} B., 1989, \aaps, 81,
  309

\bibitem[{{Ricca} {et~al.}(2012){Ricca}, {Bauschlicher}, {Boersma}, {Tielens},
  \& {Allamandola}}]{Ricca:2012aa}
{Ricca} A., {Bauschlicher} Jr. C.~W., {Boersma} C., {Tielens} A.~G.~G.~M.,
  {Allamandola} L.~J., 2012, \apj, 754, 75

\bibitem[{{Roberts} {et~al.}(2012){Roberts}, {Smith}, \&
  {Sarre}}]{2012MNRAS.421.3277R}
{Roberts} K.~R.~G., {Smith} K.~T., {Sarre} P.~J., 2012, \mnras, 421, 3277

\bibitem[{{Schutte} {et~al.}(1993){Schutte}, {Tielens}, \&
  {Allamandola}}]{Schutte:1993aa}
{Schutte} W.~A., {Tielens} A.~G.~G.~M., {Allamandola} L.~J., 1993, \apj, 415,
  397

\bibitem[{{Scott} \& {Duley}(1996)}]{Scott:1996aa}
{Scott} A., {Duley} W.~W., 1996, \apjl, 472, L123

\bibitem[{{Scott} {et~al.}(1997){Scott}, {Duley}, \& {Jahani}}]{Scott:1997ab}
{Scott} A.~D., {Duley} W.~W., {Jahani} H.~R., 1997, \apjl, 490, L175

\bibitem[{{Sellgren} {et~al.}(2010){Sellgren}, {Werner}, {Ingalls}, {Smith},
  {Carleton}, \& {Joblin}}]{2010ApJ...722L..54S}
{Sellgren} K., {Werner} M.~W., {Ingalls} J.~G., {Smith} J.~D.~T., {Carleton}
  T.~M., {Joblin} C., 2010, \apjl, 722, L54

\bibitem[{{Sharpee} {et~al.}(2007){Sharpee}, {Zhang}, {Williams}, {Pellegrini},
  {Cavagnolo}, {Baldwin}, {Phillips}, \& {Liu}}]{Sharpee:2007aa}
{Sharpee} B., {Zhang} Y., {Williams} R., {Pellegrini} E., {Cavagnolo} K.,
  {Baldwin} J.~A., {Phillips} M., {Liu} X.-W., 2007, \apj, 659, 1265

\bibitem[{{Sloan} {et~al.}(2007){Sloan}, {Jura}, {Duley}, {Kraemer},
  {Bernard-Salas}, {Forrest}, {Sargent}, {Li}, {Barry}, {Bohac}, {Watson}, \&
  {Houck}}]{Sloan:2007aa}
{Sloan} G.~C., {Jura} M., {Duley} W.~W., {Kraemer} K.~E., {Bernard-Salas} J.,
  {Forrest} W.~J., {Sargent} B., {Li} A., {Barry} D.~J., {Bohac} C.~J.,
  {Watson} D.~M., {Houck} J.~R., 2007, \apj, 664, 1144

\bibitem[{{Sloan} {et~al.}(2005){Sloan}, {Keller}, {Forrest}, {Leibensperger},
  {Sargent}, {Li}, {Najita}, {Watson}, {Brandl}, {Chen}, {Green},
  {Markwick-Kemper}, {Herter}, {D'Alessio}, {Morris}, {Barry}, {Hall}, {Myers},
  \& {Houck}}]{Sloan:2005aa}
{Sloan} G.~C., {Keller} L.~D., {Forrest} W.~J., {Leibensperger} E., {Sargent}
  B., {Li} A., {Najita} J., {Watson} D.~M., {Brandl} B.~R., {Chen} C.~H.,
  {Green} J.~D., {Markwick-Kemper} F., {Herter} T.~L., {D'Alessio} P., {Morris}
  P.~W., {Barry} D.~J., {Hall} P., {Myers} P.~C., {Houck} J.~R., 2005, \apj,
  632, 956

\bibitem[{{Speck} {et~al.}(2009){Speck}, {Corman}, {Wakeman}, {Wheeler}, \&
  {Thompson}}]{Speck_09_graphite}
{Speck} A.~K., {Corman} A.~B., {Wakeman} K., {Wheeler} C.~H., {Thompson} G.,
  2009, \apj, 691, 1202

\bibitem[{{Srinivasan} {et~al.}(2010){Srinivasan}, {Sargent}, {Matsuura},
  {Meixner}, {Kemper}, {Tielens}, {Volk}, {Speck}, {Woods}, {Gordon},
  {Marengo}, \& {Sloan}}]{Srinivasan_10_Cstars}
{Srinivasan} S., {Sargent} B.~A., {Matsuura} M., {Meixner} M., {Kemper} F.,
  {Tielens} A.~G.~G.~M., {Volk} K., {Speck} A.~K., {Woods} P.~M., {Gordon} K.,
  {Marengo} M., {Sloan} G.~C., 2010, \aap, 524, A49

\bibitem[{{Stanghellini} \& {Haywood}(2010)}]{Stanghellini:2010aa}
{Stanghellini} L., {Haywood} M., 2010, \apj, 714, 1096

\bibitem[{{Storey} \& {Hummer}(1995)}]{Storey_1995}
{Storey} P.~J., {Hummer} D.~G., 1995, \mnras, 272, 41

\bibitem[{{Tajitsu} \& {Tamura}(1998)}]{Tajitsu:1998aa}
{Tajitsu} A., {Tamura} S., 1998, \aj, 115, 1989

\bibitem[{{Vassiliadis} \& {Wood}(1994)}]{1994ApJS...92..125V}
{Vassiliadis} E., {Wood} P.~R., 1994, \apjs, 92, 125

\bibitem[{{Wang} \& {Liu}(2007)}]{2007MNRAS.381..669W}
{Wang} W., {Liu} X.-W., 2007, \mnras, 381, 669

\bibitem[{{Weidmann} \& {Gamen}(2011{\natexlab{a}})}]{Weidmann:2011ab}
{Weidmann} W.~A., {Gamen} R., 2011{\natexlab{a}}, \aap, 531, A172

\bibitem[{{Weidmann} \& {Gamen}(2011{\natexlab{b}})}]{Weidmann:2011aa}
---, 2011{\natexlab{b}}, \aap, 526, A6

\bibitem[{{Williams} {et~al.}(2008){Williams}, {Jenkins}, {Baldwin}, {Zhang},
  {Sharpee}, {Pellegrini}, \& {Phillips}}]{Williams:2008aa}
{Williams} R., {Jenkins} E.~B., {Baldwin} J.~A., {Zhang} Y., {Sharpee} B.,
  {Pellegrini} E., {Phillips} M., 2008, \apj, 677, 1100

\bibitem[{{Woods} {et~al.}(2011){Woods}, {Oliveira}, {Kemper}, {van Loon},
  {Sargent}, {Matsuura}, {Szczerba}, {Volk}, {Zijlstra}, {Sloan}, {Lagadec},
  {McDonald}, {Jones}, {Gorjian}, {Kraemer}, {Gielen}, {Meixner}, {Blum},
  {Sewi{\l}o}, {Riebel}, {Shiao}, {Chen}, {Boyer}, {Indebetouw}, {Antoniou},
  {Bernard}, {Cohen}, {Dijkstra}, {Galametz}, {Galliano}, {Gordon}, {Harris},
  {Hony}, {Hora}, {Kawamura}, {Lawton}, {Leisenring}, {Madden}, {Marengo},
  {McGuire}, {Mulia}, {O'Halloran}, {Olsen}, {Paladini}, {Paradis}, {Reach},
  {Rubin}, {Sandstrom}, {Soszy{\'n}ski}, {Speck}, {Srinivasan}, {Tielens}, {van
  Aarle}, {van Dyk}, {van Winckel}, {Vijh}, {Whitney}, \&
  {Wilkins}}]{Woods_11_classification}
{Woods} P.~M., {Oliveira} J.~M., {Kemper} F., {van Loon} J.~T., {Sargent}
  B.~A., {Matsuura} M., {Szczerba} R., {Volk} K., {Zijlstra} A.~A., {Sloan}
  G.~C., {Lagadec} E., {McDonald} I., {Jones} O., {Gorjian} V., {Kraemer}
  K.~E., {Gielen} C., {Meixner} M., {Blum} R.~D., {Sewi{\l}o} M., {Riebel} D.,
  {Shiao} B., {Chen} C.-H.~R., {Boyer} M.~L., {Indebetouw} R., {Antoniou} V.,
  {Bernard} J.-P., {Cohen} M., {Dijkstra} C., {Galametz} M., {Galliano} F.,
  {Gordon} K.~D., {Harris} J., {Hony} S., {Hora} J.~L., {Kawamura} A., {Lawton}
  B., {Leisenring} J.~M., {Madden} S., {Marengo} M., {McGuire} C., {Mulia}
  A.~J., {O'Halloran} B., {Olsen} K., {Paladini} R., {Paradis} D., {Reach}
  W.~T., {Rubin} D., {Sandstrom} K., {Soszy{\'n}ski} I., {Speck} A.~K.,
  {Srinivasan} S., {Tielens} A.~G.~G.~M., {van Aarle} E., {van Dyk} S.~D., {van
  Winckel} H., {Vijh} U.~P., {Whitney} B., {Wilkins} A.~N., 2011, \mnras, 411,
  1597

\bibitem[{{Zhang} \& {Kwok}(1993)}]{1993ApJS...88..137Z}
{Zhang} C.~Y., {Kwok} S., 1993, \apjs, 88, 137

\bibitem[{{Zhang} {et~al.}(2009){Zhang}, {Jiang}, \& {Li}}]{Zhang_09_30mic}
{Zhang} K., {Jiang} B.~W., {Li} A., 2009, \apj, 702, 680

\bibitem[{{Zhang} \& {Kwok}(2011)}]{2011ApJ...730..126Z}
{Zhang} Y., {Kwok} S., 2011, \apj, 730, 126

\bibitem[{{Zhukovska} \& {Gail}(2008)}]{Zhukovska_08_MgS}
{Zhukovska} S., {Gail} H.-P., 2008, \aap, 486, 229

\bibitem[{Zijlstra {et~al.}(2006)Zijlstra, Matsuura, Wood, Sloan, Lagadec, {van
  Loon}, Groenewegen, Feast, Menzies, Whitelock, Blommaert, Cioni, Habing,
  Hony, Loup, \& Waters}]{Zijlstra_06_LMC}
Zijlstra A.~A., Matsuura M., Wood P.~R., Sloan G.~C., Lagadec E., {van Loon}
  J.~T., Groenewegen M. A.~T., Feast M.~W., Menzies J.~W., Whitelock P.~A.,
  Blommaert J. A. D.~L., Cioni M. R.~L., Habing H.~J., Hony S., Loup C., Waters
  L. B. F.~M., 2006, \mnras, 370, 1961

\end{thebibliography}
\end{document}